%% file: main.tex
\documentclass{aa}  

\usepackage{graphicx}
 \usepackage[varg]{txfonts}
\usepackage{xcolor}
\usepackage{natbib}
 \bibpunct{(}{)}{;}{a}{}{,} 

\usepackage[angle-symbol-over-decimal, 
    range-units=single, 
    separate-uncertainty=true]{siunitx} 
\usepackage{xcolor}

\DeclareSIUnit\jansky{Jy}
\DeclareSIUnit\erg{erg}
\DeclareSIUnit\parsec{pc}
\DeclareSIUnit\msun{M_\odot}
\DeclareSIUnit\lumsol{L_\odot}
\DeclareSIUnit\zsun{Z_\odot}
\DeclareSIUnit\year{yr}
\DeclareSIUnit\beam{beam}

\usepackage{multirow}

\newcommand{\submmw}{\SI{870}{\micro\meter}}
\newcommand{\co}[2]{\ensuremath{\element[][12]{CO}\left(#1\to#2\right)}}
\newcommand{\neutralc}{\ensuremath{\left[\ion{C}{i}\right]_{1 \to 0}}}
\newcommand{\sbarc}{Sunburst Arc}
\newcommand{\planck}{PSZ1G311}

\usepackage{ulem}
\usepackage{hyperref}
\hypersetup{
    colorlinks=true,
    linkcolor=blue,
    citecolor=blue,
    filecolor=magenta,      
    urlcolor=cyan,
}

\defcitealias{Saintonge2013MolGasLensedGalaxies}{S13}
\defcitealias{Tacconi2018PHIBSS2MainSequence}{T18}
\defcitealias{DessaugesZavadsky2015MolGasLensedSFGs}{DZ15}
\defcitealias{Genzel2015DustAndCOscalingRelations}{G15}
\defcitealias{Genzel2012MetallicityCOtoH2}{G12}
\begin{document}

   \title{Molecular gas budget and characterization of intermediate-mass star-forming galaxies at $z\approx 2-3$}
   \titlerunning{Mol. gas in lensed galaxies}

   \author{M. Solimano
          \inst{1}\fnmsep\inst{3}
          \and
          J. Gonz\'alez-L\'opez\inst{2}\fnmsep\inst{1}
          \and
          L. F. Barrientos\inst{3}
          \and
          M. Aravena\inst{1}
          \and
          S. L\'opez\inst{4}
          \and
          N. Tejos\inst{5}
          \and
          K. Sharon\inst{6}
          \and
          H. Dahle\inst{7}
          \and
          M. Bayliss\inst{8}
          \and
          C. Ledoux\inst{9}
          \and
          J. R. Rigby\inst{10}
          \and
          M. Gladders\inst{11}
          }

  \institute{
            N\'ucleo de Astronom\'ia de la Facultad de Ingenier\'ia y Ciencias, %
            Universidad Diego Portales, Av.  Ej\'ercito Libertador 441, Santiago, Chile, \email{manuel.solimano@mail.udp.cl} 
         \and
            Las Campanas Observatory, Carnegie Institution of Washington, %
            Casilla 601, La Serena, Chile
        \and 
            Instituto de Astrof\'isica, Facultad de F\'isica, %
            Pontificia Universidad Cat\'olica de Chile Av.  Vicu\~na Mackenna 4860, %
            782-0436 Macul, Santiago, Chile
        \and
            Departamento de Astronom\'ia, Universidad de
            Chile, Casilla 36-D, Santiago, Chile
        \and
            Instituto de F\'isica, Pontificia Universidad Cat\'olica de Valpara\'iso, Casilla 4059, Valpara\'iso, Chile
        \and
            Department of Astronomy, University of Michigan, 1085 South University Avenue, Ann Arbor, MI 48109, USA
        \and
            Institute of Theoretical Astrophysics, University of Oslo, P.O. Box 1029, Blindern, NO-0315 Oslo, Norway
        \and
            Department of Physics, University of Cincinnati, Cincinnati, OH 45221, USA
        \and
            European Southern Observatory, Alonso de C\'ordova 3107, Vitacura, Casilla 19001, Santiago, Chile
        \and
            Observational Cosmology Lab, NASA Goddard Space Flight Center, Code 665, 8800 Greenbelt Rd, Greenbelt, MD 20771 USA
        \and
            Department of Astronomy \& Astrophysics, The University of Chicago, 5640 S. Ellis Avenue, Chicago, IL 60637, USA
        }

  \date{Received July 20, 2021; accepted August 20, 2021}


   \abstract{Star-forming galaxies (SFGs) with stellar masses below \SI{e10}{\msun} make up the bulk of the galaxy population at $z>2$. The properties of the cold gas in these galaxies can only be probed in very deep observations or by targeting strongly lensed galaxies.
   Here we report the results of a pilot survey using the Atacama Compact Array (ACA) of molecular gas in the most strongly magnified galaxies selected as giant arcs in optical data. The selection in rest-frame ultraviolet (UV) wavelengths ensures that sources are regular SFGs, without a priori indications of intense dusty starburst activity.
   We conducted Band 4 and Band 7 observations to detect mid-$J$ CO, [\ion{C}{i}] and thermal continuum as molecular gas tracers from four strongly lensed systems at $z\approx2-3$: our targets are SGAS J1226651.3+215220 (A and B), SGAS J003341.5+024217 and the \sbarc. The measured molecular mass was then projected onto the source plane with detailed lens models developed from high resolution \textit{Hubble} Space Telescope observations. Multiwavelength photometry was then used to obtain the intrinsic stellar mass and star formation rate via spectral energy distribution modeling.
   In only one of the sources are the three tracers robustly detected, while in the others they are either undetected or detected in continuum only. The implied molecular gass masses range from \SI{4e9}{\msun} in the detected source to an upper limit of $\lesssim\SI{e9}{\msun}$ in the most magnified source. The inferred gas fraction and gas depletion timescale are found to lie approximately $0.5$ to $1.0$ dex below the established scaling relations based on previous studies of unlensed massive galaxies, but in relative agreement with existing literature about UV-bright lensed galaxies at these high redshifts.
   Our results indicate that the cold gas content of intermediate to low mass galaxies should not be extrapolated from the trends seen in more massive high-$z$ galaxies. The apparent gas deficit is robust against biases in the stellar mass or star formation rate. However, we find that in this mass-metallicity range, the molecular gas mass measurements are severely limited by uncertainties in the current tracer-to-gas calibrations. }
   

   \maketitle
%

\section{Introduction}\label{sec:intro}

New stars are born from giant, cold molecular gas clouds in dusty environments within the interstellar medium (ISM) of star-forming galaxies \citep[SFGs,][]{ Bolatto2008ExtragalacticGMCs, KennicuttAndEvans2012ReviewSFLocalUniverse}, although the detailed physical mechanisms by which galaxies acquire gas and then convert it into stars are still poorly understood \citep[e.g.,][]{McKeeAndOstriker2007TheoryOfSF}. However, theoretical and observational efforts in the past decades have contributed to linking the evolution of galactic-scale parameters of star formation (e.g., star formation rate, SFR; stellar mass, M\textsubscript{stars}; gas depletion timescale, $t_\text{dep}=\text{M}_\text{mol}/\text{SFR}$; etc.) with the abundance and physical conditions of the cold ISM from the Local Universe up to redshift $z\sim4$ \citep[see the recent reviews by][]{Hodge&daCunha2020Review, Tacconi2020EvolutionISMReview, ForsterShreiberAndWuyts2020CosmicNoonSFReview}.

The advent of large samples of massive SFGs with targeted dust continuum and/or CO line observations (the two most common molecular gas tracers at high redshift) revealed a number of key trends: a tight correlation between the gas fraction (M\textsubscript{mol}/M\textsubscript{stars}) and the specific star formation rate (sSFR$\equiv \text{SFR}/\text{M}_\text{stars}$), a decrease in $t_\text{dep}$ with redshift, and an increase in the gas fraction with redshift \citep[e.g.,][]{Scoville2017EvolutionISM, Tacconi2018PHIBSS2MainSequence}. These results imply that galaxies with higher levels of star formation also have larger molecular gas reservoirs, and that these reservoirs were on average larger in the first gigayears of the universe. The current explanation is that the ability of a galaxy to form stars is regulated by a feedback loop between the cosmological accretion of gas and the return of that gas to the ambient via galactic outflows, a picture that is often refered to as the "equilibrium model" \citep{Dave2012EquilibriumModelAnalytical, Lilly2013GasRegulation, Walter2020EvolutionOfBaryons}.  Furthermore, deep blind \citep[ASPECS,][]{Decarli2019ASPECS-CO-LF, Decarli2020ASPECSMultiBandLineLuminosityAndGasDensity} and archival \citep[A\textsuperscript{3}COSMOS, ][]{Liu2019A3COSMOScoldGasEvolution} surveys conducted with the Atacama Large Millimeter/Submillimeter Array (ALMA) confirm that the cosmic molecular gas density follows roughly the same evolution as the cosmic star formation density, that is with a global maximum occurring around $z\approx 2$ \citep[e.g.,][]{Madau&Dickinson2014Review}. However, at $z\gtrsim1$ this picture is necessarily incomplete, since the current statistical samples are limited to the highest-mass systems ($\SI{e10}{\msun} \leq M_\text{stars} \leq \SI{e11.5}{\msun}$) despite lower mass galaxies being more numerous. Unfortunately, directly probing the gas budget of the $<\SI{e10}{\msun}$ galaxies at $z\sim 2$ is still a major observational challenge, even with the exquisite sensitivity of ALMA, and has so far been achieved only through stacking techniques \citep{Inami20202StacksAspecsCO}. Low-mass galaxies tend to have a low oxygen abundance (hereafter metallicity) as expected from the mass-metallicity relation \citep[MZR; e.g.,][]{Tremonti2004MassMetallicityRelation}. This makes the detection of dust and CO even more difficult, since dust is less abundant in low-metallicity environments, favoring the photodissociation of CO molecules \citep{Leroy2011COtoH2, Genzel2012MetallicityCOtoH2, Bolatto2013COtoH2}.

A popular alternative for exploring the low-mass regime at high-redshift is to take advantage of strong gravitational lensing.
This phenomenon occurs when a distant galaxy is closely aligned along the line of sight with a foreground massive galaxy or cluster of galaxies. The space-time deflection induced by the intervening mass -- the lens -- produces distorted, magnified, and in some cases, multiple images of the background galaxy projected in the sky, often forming rings or giant arcs. The lensed images appear brighter than they would in the absence of this effect, providing a boost in intrinsic flux limit, that is, at a given instrumental sensitivity, lensing enables the detection of intrinsically fainter sources. Since the lensing effect is stronger toward massive galaxy clusters, the brightest giant arcs are typically found near their cores. Thus, targeting cluster-lensed galaxies became a common strategy for pushing molecular gas studies to fainter and higher redshift galaxies, which provides reliable results if a good model of the lens deflection is available \citep[e.g., ][]{Smail1997SubmmSurveyOfLensingClusters, Baker2004MolGasInCB58, Danielson2011StarformingGalaxyAtRedshift2, Saintonge2013MolGasLensedGalaxies, Sharon2013MappingCOinLensedSMG, DessaugesZavadsky2015MolGasLensedSFGs, Spingola2020SHARP6ExtendedCOLensed}. 

Although giant arcs are rare, directed searches in wide field optical imaging surveys have successfully identified several dozens to this date \citep{Hennawi2008SGAS1, Kubo2010SGAS, Stark2013CassowaryLensedSample, Khullar2021COOLLAMPSBrightestLensedGalaxy}. This number is expected to grow substantially with  upcoming surveys such as the Rubin Observatory Legacy Survey of Space and Time \citep[LSST;][]{Ivezic2019LSST}, the Euclid mission \citep{Laureijs2011EuclidDefinition}, or the Roman Space Telescope \citep{Spergel2015RomanReport}. For example, Euclid is expected to detect $\sim 2\,000$ giant arcs in galaxy clusters with a contrast of at least $3\sigma$ above the background \citep{Boldrin2012PredictedGiantArcsNumber}. The detection of lensed systems in broadband rest-frame UV imaging ensures that the sources are not heavily dust-obscured and can be selected as SFGs by their photometric colors. In contrast to the massive and dust-rich submillimeter-selected galaxies (SMGs), sources selected by their colors represent the bulk of the SFG population at $z=2$, which has moderate dust and gas content.

The study of cold gas in UV-selected lensed SFGs dates back to the pre-ALMA era. \citet{Saintonge2013MolGasLensedGalaxies} \citepalias[hereafter][]{Saintonge2013MolGasLensedGalaxies} obtained far infrared photometry (using the \textit{Herschel} Space Telescope) and  CO line measurements for 17 lensed galaxies selected from the literature as optically-bright giant gravitational arcs. These galaxies lie in the star-forming main sequence (MS) with stellar masses in the \SIrange{e9.7}{e10.8}{\msun} range. Their analysis revealed lower depletion timescales and similar gas fractions than $z \sim 1$ MS galaxies, suggesting a flattening in the trend of increasing gas fraction with redshift. They also provide evidence for redshift evolution of the gas-to-dust ratio, a quantity that is commonly assumed to be constant. \citet{DessaugesZavadsky2015MolGasLensedSFGs} \citepalias[hereafter][]{DessaugesZavadsky2015MolGasLensedSFGs} found similar results by combining five additional lensed galaxies with larger samples of SFGs at $z > 1$. The inclusion of galaxies with even lower stellar masses (down to \SI{e9.4}{\msun}), clearly showed the effect of  ``downsizing'' in the gas fraction versus redshift relation: the increase in the gas fraction with redshift is more pronounced for low M\textsubscript{stars} systems.

Both studies are based on unresolved observations of the cold ISM. With ALMA, it is now possible to exploit the enhanced resolving power delivered by gravitational lensing to characterize the molecular gas and dust at sub-kiloparsec scales \citep[e.g.,][]{ALMAPartnership2015LensedSMGz3}. This has allowed to gain unprecedented insight about the kinematics, sizes, distribution and CO excitation of individual star-forming clumps. Recent work has revealed that the high redshift ISM has more compact dusty cores, increased turbulence, higher densities relative to local SFGs, together with deviations from the Schmidt-Kennicutt law \citep[e.g.,][]{Messias2014HATLAS-J1429Merger, Swinbank2015ResolvedALMADiskAtz3,Canameras2017DustyGemsResolved60pc, Apostolovski2019ImagingMolGasSPTatz5.7, Rybak2020FullOfOrions, Rizzo2020ColdDiskSPT}. However, the majority of the high-resolution studies still focus on the most massive dusty systems and not on the  more common low mass, low metallicity galaxies. The reason for this is that strong lensing multiplies intrinsic area by a factor $\mu$, but surface brightness remains constant. While this effectively boosts flux by the same factor $\mu$,  resolving the emission requires significant integration time regardless of lensing. Interferometric follow-up campaigns on UV-selected, intermediate- to low-mass lensed SFGs have faced mixed success: on one side, some studies detect dust continuum and CO line emission at high significance, enabling an state-of-the-art analysis of the properties of the cold ISM and their connection to star formation within very small physical scales (\citealt{DessaugesZavadsky2017MolGasLensedCaseStudy, Gonzalez-Lopez2017LensedMolGasInRCS0327}; in the case of the "Cosmic Snake", down to $\sim \SI{30}{\parsec}$ at $z=1$, \citealt{Dessauges-Zavadsky2019CosmicSnake}). On the other side, some studies can only provide upper limits on gas and dust mass, due to the nondetection of the lensed galaxy \citep[e.g.,][]{Livermore2012NonDetectionGasRedshift5, Rybak2021UltrafaintLensedCII} 

Even with the help of lensing, measuring the cold ISM in high-$z$ dwarf galaxies is  challenging because the high magnifications do not guarantee detection. For this reason, we have launched a low-resolution exploratory campaign to detect mid-J CO line emission and dust continuum in the optically brightest giant arcs known, to systematically expand the number of lensed low-mass SFGs with measured cold ISM properties.

In this paper we present the first results from this program, where we conducted low resolution interferometric observations of four extremely magnified systems at $z \sim 2.5$ with the Atacama Compact Array (ACA).  The detections are used to locate the intrinsic properties of the galaxies within the scaling relations found in previous works and also as a starting point for future follow-up campaigns with extended ALMA configurations.

The paper is organized as follows. In Sect. \ref{sec:obs} we describe the sources selected for this study , the observing setup,  and the ancillary dataset. In Sect. \ref{sec:results} we explain our methods and present our main results. Finally, the interpretation and discussion of the results is given in Sect. \ref{sec:discussion}.

We adopt a flat $\Lambda$CDM cosmology with $h=0.7$, $\Omega_m = 0.3$ and $\Omega_\Lambda = 0.7$. All position angles are quoted east of north and all magnitudes are given in the AB system. Star formation rates and stellar masses are derived assuming a \citet{Chabrier2003IMF} initial mass function (IMF). Also, relative velocities follow the radio convention (i.e., linear approximation in frequency, rather than wavelength).

\section{Observations} \label{sec:obs}

\subsection{Sample selection and prior identification} \label{sec:selection}

We targeted four giant gravitational arcs selected for their extreme brightness in optical bands ($m_{V} < 21$). The first one, nicknamed the \sbarc, was discovered in an optical follow-up of the Planck Sunyaev-Zeldovich cluster candidate \object{PSZ1 G311.65-18.48} (hereafter \planck) and identified as a SFG at $z=2.37$ \citep{Dahle2016SunburstArcDiscovery}. Since discovery, it holds the title of the brightest lensed image of a galaxy known to this date, and has been confirmed to be leaking significant amounts of ionizing photons \citep{Rivera-Thorsen2017SunburstArcLymanEscape, Rivera-Thorsen2019SunburstArcEscapeFractions}. The galaxy is lensed into multiple images, grouped in four distinct arc segments. Hereafter, we refer to each segment S1, S2, S3 as in the discovery paper (see Fig. \ref{fig:rb7_continua}). The second and third targets are the two components of \object{SGAS J1226651.3+215220} (hereafter SGASJ1226) at $z=2.92$, which were discovered in the Sloan Digital Sky Survey \citep[SDSS,][]{Blanton2017SloanIV} imaging data as a part of the Sloan Giant Arc Survey \citep[SGAS;][]{Hennawi2008SGAS1} using the Lyman Break technique \citep{Koester2010ArcJ1226Discovery}. The brightest arc is the two-fold almost symmetric pair of images \citep[see][]{Dai2020AsymmetricBrightnessJ1226Arc} we refer to as SGASJ1226-A.1 (see Fig. \ref{fig:rb7_continua}). An additional but less magnified image of the same galaxy (SGASJ1226-A.2) appears roughly \ang{;;15} to the east. We call the source galaxy producing these images SGASJ1226-A. The southern, optically fainter arc is the second component of SGASJ1226, identified as companion galaxy at the same redshift \citep{Koester2010ArcJ1226Discovery}, hereafter labeled SGASJ1226-B.1 (SGASJ1226-B in the source plane). Since A and B lie within \ang{;;15} in the image plane, a single ACA pointing of the lensing cluster core was needed to cover both targets. The fourth target, \object{SGAS J003341.5+024217} (hereafter SGASJ0033) at $z=2.39$, was also discovered in SDSS data and it was first reported in \citet{Rigby2018MegaSauraI}. The source galaxy SGASJ0033-A is lensed into four different images: the first two are blended in a \ang{;;5}-long bright arc we call SGASJ0033-A.1. The remaining less magnified images are SGASJ0033-A.2 and SGASJ0033-A.3 (see Fig. \ref{fig:rb7_continua}).

Both SGASJ0033-A and SGASJ1226-A are part of the Magellan Evolution of Galaxies Spectroscopic and Ultraviolet Reference Atlas \citep[MegaSaura;][]{Rigby2018MegaSauraI}, a catalog of moderate resolution rest-frame ultraviolet spectra of 15 gravitationally lensed galaxies at $z\sim 2$ taken with the MagE instrument on the \textit{Magellan} telescopes. The \sbarc~has comparable MagE data as part of the Extended MegaSaura Survey (Rigby et al. in prep.). The MegaSaura spectra provide secure spectroscopic redshifts as well as robust estimates on the age and metallicities of the young stellar populations \citep{Chisholm2019MegaSauraMetallicity}. Unfortunately, a MagE spectrum for SGASJ1226-B is not available. In what follows, we adopt the stellar metallicity of SGASJ1226-B to be the same of SGASJ1226-A.
 
Promisingly, the \textit{James Webb} Space Telescope (JWST) will target two of the sources presented here, namely SGASJ1226-A (ERS 1355, PIs: Rigby \& Vieira) and the \sbarc~(GO 2555, PI: Rivera-Thorsen). The characterization we give here of the cold ISM in these galaxies will provide crucial insight to interpret the upcoming data.

\begin{table*}[!hbt]
    \centering
    \caption{\label{tab:summary_targets} Summary of selected target and ACA observations.}
    \input{summary_targets}
    \tablebib{
    (1)~\citet{Rivera-Thorsen2017SunburstArcLymanEscape};  (2)~\citet{Rigby2018MegaSauraI}
    }
    \tablefoot{Coordinates indicate the phase center of the ACA pointing. 
    \tablefoottext{a}{Root mean square computed in line-free regions of the zeroth moment residual map.}
    \tablefoottext{b}{Root mean square computed in the dirty image with the sources (if any) masked out.}
    \tablefoottext{c}{Computed as the square root of the product of the major and minor axis length of the beam.}
    }
\end{table*}

\subsection{Atacama Compact Array data} \label{sec:aca_data}
Our data includes observations with ACA obtained under ALMA program 2018.1.01142.S (PI: Gonz\'alez-L\'opez). ACA -- also known as the Morita Array -- is a fixed configuration of twelve 7-meter antennas with a maximum baseline of 48.9 meters located at the Llano of Chajnantor \citep{Iguchi2009AtacamaCompactArray}. Our sources are lensed arcs that extend over several arcseconds on the sky, and ACA offers an optimal combination of an angular resolution that is high enough to deblend the sources while remaining sensible to large scale emission. 

For all targets we set up observations to detect mid-$J$ \element[][12]{CO} emission lines in ALMA Band 4 ($\nu\sim\SI{140}{\giga\hertz}$) and continuum dust emission in ALMA Band 7 ($\nu\sim\SI{350}{\giga\hertz}$). Mid-$J$ transitions have been reported to be one of the brightest for most of star-forming galaxies \citep{Carilli&Walter2013Review, Daddi2015COexciation, Boogaard2020ASPECS-COExcitation} while also providing information about the molecular gas budget. Band 7 observations, on the other hand,  are regularly used to trace dust-obscured star formation and also molecular gas mass \citep[e.g.,][]{Schinnerer2016GasFractionAndDepletion,Scoville2017EvolutionISM,Miettinen2017AlmaCosmosSED,Darvish2018GasScalingRelations,Casey2019Mambo9,Magnelli2020ASPECSDustAndGasDensity}. In addition, Band 7 photometry for sources at $z\sim 2.5$ puts important constraints on their overall far-infrared (FIR) luminosity.

All ACA data were calibrated and reduced using the Common Astronomy Software Application\footnote{\url{https://casa.nrao.edu/}} \citep[CASA;][]{McMullin2007CASA} version 5.4.0-70 following the standard pipeline scripts. We assumed a flux calibration accuracy of 10\% throughout. All subsequent imaging was done with CASA task \verb|tclean| and using natural weights to maximize signal-to-noise ratio. The resulting cubes and images are sampled at pixel size equal to 1/5 of beam minor axis, and the astrometry is expected to be accurate to \ang{;;0.6}.

\subsubsection{The \sbarc}
We used a standard continuum setup for observing the \sbarc~with Band 7 receivers. This comprises a total of four spectral windows with \SI{31.250}{\mega\hertz} wide channels. Two adjacent spectral windows cover from \SIrange{335.504}{339.426}{\giga\hertz} and the other two from \SIrange{347.504}{351.483}{\giga\hertz}.

From the observed visibilities, we performed dirty imaging in multifrequency synthesis mode. The process yielded a synthesized beam of $\ang{;;5.1}\times\ang{;;4.8}$ with position angle (PA) of \ang{5.6;;}. The resulting continuum intensity map achieved a root-mean-square (RMS) sensitivity of $1\sigma \simeq \SI{245}{\micro\jansky}$ over a total bandwidth of \SI{7.5}{\giga\hertz} and 145 minutes on source. We found no peaks higher than $3\sigma$ within the primary beam (\ang{;;43} in diameter), accounting for a nondetection of the arc in this experiment (see Fig. \ref{fig:rb7_continua}) 

The spectral configuration for Band 4 included two main spectral windows. The first one was tuned to the frequency of the $\co{4}{3}$ transition at $z=2.37$, covering from \SIrange{134.472}{137.469}{\giga\hertz} at a channel
resolution of \SI{3.904}{\mega\hertz} (equivalent to \SI{8.5}{\kilo\meter\per\second} at the CO central frequency). The other one
targeted the $\neutralc$ atomic line by covering from
\SIrange{145.553}{147.526}{\giga\hertz} at the same resolution (\SI{8}{\kilo\meter\per\second}). The setup also
included two adjacent sidebands to constrain the underlying continuum emission. 

We then synthesized data-cubes keeping the native channel resolution (\SI{3.9}{\mega\hertz}). We reached a sensitivity of $1\sigma\simeq\SI{2.1}{\milli\jansky\per\beam}$ per channel and a median synthesized beam of $\ang{;;11.74} \times \ang{;;11.15}$ (PA=\ang{-17;;}) for the $\co{4}{3}$ spectral window and $\ang{;;10.85}\times\ang{;;10.35}$ (PA=\ang{-17;;})
for the $\neutralc$ spectral window. We found that the cube contains no evident emission line signatures, that is, none of the targeted lines were detected in this experiment. However, we found a tentative ($\lesssim 3\sigma$) detection of a continuum source over the S1 arc segment (see Fig. \ref{fig:rb7_continua}) after averaging all the channels, including the sidebands. We then repeated the imaging process excluding channels within \SI{250}{\kilo\meter\per\second} to each side of the expected central frequency for the $\co{4}{3}$ and $\neutralc$ lines. To clean out the dirty beam we manually placed a \ang{;;10} mask over the peak and cleaned down to the $1\sigma$ level. The final image has a  beam of $\ang{;;11.5} \times \ang{;;10.7}$ (PA=-\ang{15.9}) and residual RMS of \SI{80.4}{\micro\jansky\per\beam} achieved with 226 minutes of on-source integration time. 

\subsubsection{SGASJ1226}\label{sec:j1226_aca_description}
For the SGASJ1226 system we used the same spectral setup used for Band 7 continuum detection in the \sbarc, with the same coverage in frequency and spectral resolution. The synthesis of the dirty image revealed two sources with signal-to-noise ratio greater than five: A $5.5\sigma$ peak that is cospatial with SGASJ1226-B.1 (southward of the BCG, Fig. \ref{fig:rb7_continua}), and a $6.2\sigma$ peak which we associate with a known \ion{Mg}{ii} absorber galaxy at $z=0.77$ \citep[][therein labeled SGASJ1226-G1]{Mortensen2021ScoopArctomoJ1226, Tejos2021ArctomoJ1226Kinematic}. We cleaned the images by placing clean masks on top of the two detected sources and iterating down to $2\sigma$ level. The cleaned image allowed to resolve both sources unambiguously with a synthesized beam size of $\ang{;;4.9}\times\ang{;;3.8}$ (PA=-\ang{57}) and RMS of $1\sigma \simeq \SI{170}{\micro\jansky\per\beam}$ obtained with 202 minutes of on-source integration time (see Fig. \ref{fig:rb7_continua}).

 Due to the higher redshift of the SGASJ1226 source ($z=2.92$), we used Band 4 receivers to observe only the $\co{5}{4}$ transition, as no other strong lines are expected at this redshift. Thus the setup covers with \SI{3.9}{\mega\hertz} resolution ($\sim\SI{8}{\kilo\meter\per\second}$) from
 \SIrange{145.034}{146.905}{\giga\hertz} and from
 \SIrange{146.894}{148.765}{\giga\hertz}, with an effective overlap of three channels. The corresponding sidebands cover from \SIrange{132.906}{134.890}{\giga\hertz} and \SIrange{134.905}{136.890}{\giga\hertz} at a channel width of \SI{31.25}{\mega\hertz}.  We constructed the dirty cube using \verb|tclean| in cube mode and using natural  weighting of the visibilities. We get a median beam of $\ang{;;9.9} \times \ang{;;9.0}$ (PA=-\ang{70}) and a sensitivity of $1\sigma \simeq \SI{3.3}{\milli\jansky\per\beam}$ per channel using a total of 145.3 minutes of on-source integration time. However the result showed no significant emission anywhere in the cube, nor in the channel-averaged image.
 
 \subsubsection{SGASJ0033}\label{sec:j0033_aca_description}
 For SGASJ003341 we used again the standard Band 7 continuum setup configuration. The dirty image showed three bright sources within the primary beam (see Fig. \ref{fig:rb7_continua}). The first two are spatially coincident with the optical arc and a strong \ion{Mg}{ii} absorber at $z=1.17$ (hereafter labeled SGASJ0033-G1, Ledoux et al., in prep.), respectively. The third one is a previously unreported, very bright ($\gtrsim \SI{5}{\milli\jansky} $) source with no evident optical counterpart. For this reason we refer to it as SGASJ0033-SMG (sub-millimeter galaxy). A tentative counterpart on that position is seen only in the near infrared WFC3/F140W image of the \textit{Hubble} Space Telescope  (HST; as shown in the inset of the third panel of Fig. \ref{fig:rb7_continua}), and not in the bluer filters. The lack of an optical detection further supports the idea that the object is a high-redshift dusty SFG.  Then we cleaned the image to the 2$\sigma$ level taking advantage of the automatic masking feature of \verb|tclean|. With a total of 203 minutes of on-source integration time, the resulting image reached a continuum RMS of \SI{0.17}{\milli\jansky\per\beam}, with a beam size of $\ang{;;4.5} \times \ang{;;2.9}$ (PA=-\ang{89.7;;}).
 
Band 4  spectral configuration is similar to the one used for \planck. We targeted the $\co{4}{3}$ line (expected to lie at \SI{136}{\giga\hertz} at $z=\num{2.39}$)
on the first spectral window and the $\neutralc$ line (\SI{145.180}{\giga\hertz}
at $z=\num{2.39}$) on the second. The channel bandwidth of these spectral windows was \SI{3.906}{\mega\hertz} or \SI{8.6}{\kilo\meter\per\second} at \SI{136}{\giga\hertz}. After flagging the channels were we expect the CO and [\ion{C}{i}] lines to be, we performed first order continuum subtraction in the $uv$-plane using the CASA task \verb|uvcontsub|, since this source is also a bright continuum emitter. We then carried out the imaging process with \verb|tclean| using natural weighting and manually defined clean region boxes. With 202 minutes of on-source integration time, we achieve a sensitivity of  $1\sigma \simeq \SI{2.1}{\milli\jansky\per\beam}$ per channel 
and a median synthesized beam size of $\ang{;;13.1}\times\ang{;;6.9}$ at a
PA of \ang{-77.6;;}. Preliminary analysis revealed a $10\sigma$ detection of the CO line and a $4\sigma$ detection of [\ion{C}{i}], both cospatial with the optical arc. We also report the detection of a broad ($\sim \SI{400}{\kilo\meter\per\second}$) emission line at $\nu=\SI{135.6}{\giga\hertz}$, roughly cospatial with SGASJ0033-SMG (see lower panels of Fig. \ref{fig:coline0033}).

\begin{table}[!hbt]
    \centering
    \caption{\label{tab:continuum_flux} ACA continuum photometry in bands 7 and 4.}
    \input{table870um_detections}
    \tablefoot{%
    Labels for the targeted sources in the first column are defined in Sec. \ref{sec:selection}, whereas the names for the nontarget sources are introduced in subsections \ref{sec:j1226_aca_description} and \ref{sec:j0033_aca_description}. All quantities are quoted in the image plane, i.e., not corrected for lensing magnification. Upper limits correspond to three times the RMS divided by the primary beam attenuation factor at the source position.
    \tablefoottext{a}{The reported position is the center of the ACA pointing}
    \tablefoottext{b}{Flux from blended images SGASJ0033-A.1 and SGASJ003-A.2.}
    }
\end{table}

\subsection{Ancillary data} \label{sec:ancillary}
\subsubsection{HST data}
The three lensing clusters have been observed with the HST at multiple wide band filters. We retrieved the available data from the Mikulski Archive for Space Telescopes\footnote{\url{https://archive.stsci.edu/}} (MAST) to perform photometric extraction of each source in order to constrain their spectral energy distribution (SED). The \planck~field was observed in bands  F606W and F160W with the Wide Field Camera 3 (WFC3; GO15377, PI: Bayliss), and the F814W filter with the Advanced Camera for Surveys (ACS; GO15101, PI: Dahle). For SGASJ1226, bands F110W and F160W of the near infrared channel of WFC3 were available (GO15378, PI: Bayliss), as well as bands F606W and F814W  with the ACS (GO12368, PI: Morris). Finally, data for SGASJ0033 comprises bands F555W, F814W, F105W and F140W taken with the WFC3 (GO14170, PI: Wuyts). All imaging data were aligned with  \textsc{DrizzlePac} routine \verb|tweakreg|, and drizzled to a common pixel size of \ang{;;0.03} with \verb|astrodrizzle| using Gaussian kernels with drop size of \num{0.8}. We performed additional cosmic ray removal in \sbarc's F606W drizzled frame using \textsc{AstroScrappy} \citep{McCully2018AstroScrappy}. As a last step, we check the absolute astrometric accuracy of our WCS solution by comparing the centroid positon of foreground stars in the drizzled images with the Gaia DR2 catalog \citep{GaiaCollaboration2018Release2}. We find no systematic offset nor rotation, and the observed shift is at most \ang{;;0.1}.

Since the lensed arcs are very extended in the sky and show complex morphologies, standard extraction routines are not well suited for obtaining reliable photometry of the sources. Instead, we employed ad-hoc polygonal apertures defined in the filter with the largest PSF for each target. We note that the background in SGASJ1226 and SGASJ0033 is dominated by the light from the brightest cluster galaxy (BCG) or other bright members of the foreground cluster. For this reason, we modeled the two elliptical isophotes of the BCG that encompass the arc extent, to isolate an area with similar contamination and background level as the source. Then we randomly placed two thousand 10 pixel wide apertures between these two isophotes and we take the resulting flux mean and variance as the background level and noise variance respectively for the science aperture, after properly scaling for the aperture area. For the \sbarc, we used rectangular apertures enclosing each of the main arc segments. We estimated the background in nearby sky regions to each side of the arc.

\subsubsection{\textit{Spitzer}}
Observing programs \#70154 (PI: Gladders) and \#13111 (PI: Dahle) of the \textit{Spitzer} Space Telescope include observations of SGASJ1226 and the \sbarc~respectively, at \SI{3.6}{\micro\meter} and \SI{4.5}{\micro\meter} wavelengths using the Infrared Array Camera (IRAC). At $z\sim3$ these bands trace the older stellar populations of the source galaxies and hence are very important for constraining their total stellar mass. We retrieve the IRAC mosaics from the NASA/IPAC Infrared Science Archive reduced with the stock calibration pipeline. 

Photometry for SGASJ1226-A.1 was obtained following the method outlined in \citet{Saintonge2013MolGasLensedGalaxies}: using curve of growth arguments, we determine the optimal sized circular aperture and background annulus for computing the flux of SGASJ1226-A.1. The measured flux is \SI{54+-11}{\micro\jansky} and \SI{47+-9}{\micro\jansky} at \SI{3.6}{\micro\meter} and \SI{4.5}{\micro\meter} respectively. These values agree with the ones reported by \citet{Saintonge2013MolGasLensedGalaxies}, \SI{52.8+-6.0}{\micro\jansky} and \SI{47.2+-4.2}{\micro\jansky}. The extraction of the southern arc, SGASJ1226-B, required a segment of an annulus instead of a circular aperture due to its larger extension. The annulus was centered on the core of the BCG and the complementary segment was used for background subtraction. The resulting image plane fluxes are  \SI{46+-13}{\micro\jansky} at \SI{3.6}{\micro\meter} and \SI{44+-8}{\micro\jansky} at \SI{4.5}{\micro\meter}.

Unfortunately, the \sbarc~field suffers from severe crowding, and the arc appears blended with foreground galaxies and stars at the IRAC resolution. Thus, aperture photometry was not an option in this case. To avoid the introduction of additional uncertainties and assumptions, we did not attempt a prior-based PSF-fitting approach, but completely exclude these two bands from the analysis.

The effect on the determination of stellar mass when the IRAC bands are not included in the SED fitting was explored in \citet{Mitchell2013StellarMassSystematics} by analyzing mock galaxy SEDs with known physical parameters. In that study, the exclusion of IRAC bands resulted in a bias in M\textsubscript{stars} of $-0.04$ and $-0.03$ dex at $z=2$ and $z=3$, respectively, after removing the contribution from dust and metallicity effects. The scatter in the recovered M\textsubscript{stars} is, however, on the order of 0.1 dex. Although these values were obtained in the idealized situation where the SED fitting code matches the IMF and stellar population model used for generating the SED, we can interpret them as an indication that our M\textsubscript{stars} estimate for SGASJ0033-A  and the \sbarc~has larger uncertainties but not a significant bias. As a complementary test, we ran \textsc{Magphys} on SGASJ1226-A and SGASJ1226-B again but turning off the fitting for the IRAC bands. The median value of the marginal posterior probability distribution for M\textsubscript{stars} was $+0.02$ and $-0.11$ dex for SGASJ1226-A and SGASJ1226-B, respectively, relative to the fiducial values listed in Table \ref{tab:sed_results}. In both cases, the offset is much smaller than the uncertainty returned by the code. This again indicates that the lack of IRAC bands only affects the uncertainty but not the nominal value.

\subsubsection{Lens models and source plane reconstruction} \label{sec:lensmodels}
Lens models for the three clusters were developed with the parametric \textsc{LensTool} \citep{Jullo&Kneib2009LensTool} software based on identification of multiple images of the same lensed features in the HST imaging data, following the same methodology detailed in \citet{Sharon2020ClusterLensModelsSGAS}. The models consist of a collection of parametric dark matter halo profiles centered on the cluster galaxies and assuming all members lie on the same plane (thin lens approximation).  Once a set of parameters that minimize the spatial offset between predicted and observed image pairs is obtained, the best fit model provides the deflection matrices and magnification maps that are used to reconstruct the source plane properties of the lensed galaxies. The models for the three lensing clusters in this work were already published in the literature and all of them were fit with at least 10 lensing constraints, including features in the giant arcs as well as other multiply-imaged sources at different redshifts. We refer the reader to the following references for details on each model: the lens model for the \planck~cluster is published in \citet{Rivera-Thorsen2019SunburstArcEscapeFractions}; for the SGASJ1226 cluster, in \citet{Dai2020AsymmetricBrightnessJ1226Arc} and \citet{Tejos2021ArctomoJ1226Kinematic}; and for SGASJ0033, in \citet{Fischer2019outflowJ0033}.

Throughout this paper, we report average magnifications computed as the area ratio between the image plane and the source plane within a given aperture defined in the image plane. Whenever the aperture includes  multiple images, the magnification factor $\mu$ accounts for this to match the source plane area of the image with the largest footprint. For example, the arc SGASJ1226-A.1 is divided into two images by the lensing critical line. Each half maps to the same (partial) region of the source galaxy. SGASJ1226-A.2, on the other hand, is a less magnified version of the whole source galaxy, as it has a larger footprint in the source plane. Then, we set the magnification factor so that the joint image plane area of both halves of SGASJ1226-A.1 divided by $\mu$ yielded the source plane area of SGASJ1226-A.2.  We adopted an uncertainty on magnification of 20\% for all images to account for the statistical error in cluster lens modeling. We note, however, that the uncertainties are possibly dominated by systematics \citep[e.g.,][]{Raney2020FrontierFieldsLensingUncertainties}. 

An advantage of observing this kind of systems with millimeter and submillimeter interferometers is that the model to be used is developed from independent data. Models created with deep optical images can be used to interpret the observed emission, avoiding the use of the same data to characterize the models. This approach has proven to be the only reliable method for dealing with cluster-scale strong lensing \citep[e.g.,][]{Saintonge2013MolGasLensedGalaxies,Gonzalez-Lopez2017LensedMolGasInRCS0327, Laporte2017ALMA-FFphotometry1.1mm, Sharon2019MolGasLensedJ0901, Dessauges-Zavadsky2019CosmicSnake}. However, it is important to acknowledge that several studies of galaxy-scale systems -- which have a single or very few lensing halos -- have successfully used the interferometric data alone to fit lens models, provided such data reaches high angular resolution \citep[e.g.,][]{Hezaveh2013ALMA-SPT, Bussmann2013SMALensedz1.5,Messias2014HATLAS-J1429Merger, Rybak2015SDP.81Continuum, Rybak2015SDP.81EmissionLine,Spilker2016ALMALensModelsForSPT,Apostolovski2019ImagingMolGasSPTatz5.7,Rizzo2020ColdDiskSPT}

\section{Data analysis \& results} \label{sec:results}
\subsection{Emission lines in the SGASJ0033 system} \label{sec:emission_lines}
\begin{figure}[!htb]
    \centering
    \resizebox{\hsize}{!}{\includegraphics{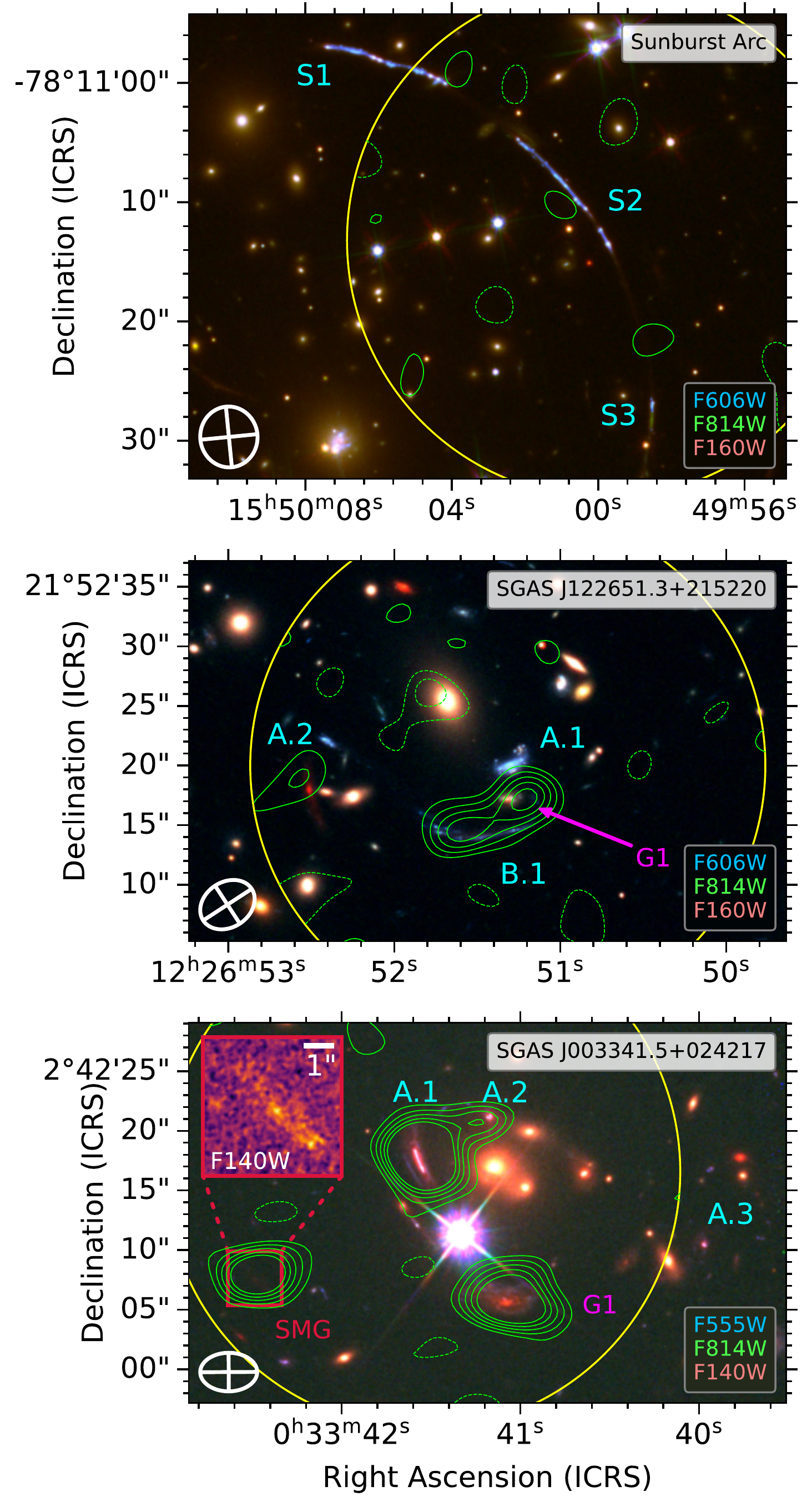}}
    \caption{Band 7 continuum contours for the three ACA pointings in this study on top of their respective HST color images. In all three panels, green solid (dashed) lines indicate positive (negative) contours of \submmw~imaging at the $\pm 2,3,4,5,6\sigma$ levels. The yellow circle indicates the 20\% sensitivity cut-off of the antenna primary beam. The white crossed ellipses at the lower left corners show the size of the synthesized beams. \textit{Upper panel:} No significant detections are obtained within ACA data. \textit{Middle panel:} Two blobs of unambiguous emission are detected toward the center of the ACA beam, roughly cospatial with G1 and B1. \textit{Lower panel:} Four bright sources are detected at $\lambda=\submmw$. Three of them are associated with A.1, A.2 and G1 respectively whereas the fourth source to the east is a previously unreported source (SMG). The inset shows a close-up view of a possible counterpart of the SMG in the WFC3-IR-F140W image (see Sect. \ref{sec:j0033_aca_description}).}
    \label{fig:rb7_continua}
\end{figure}

To search for lower S/N emission lines within the ACA cubes, we conducted two experiments: first, we selected and integrated along the velocity axis the channels where lines are expected according to redshift priors from literature \citep[e.g.,][]{Koester2010ArcJ1226Discovery, Tejos2021ArctomoJ1226Kinematic} using a fixed width of \SI{200}{\kilo\meter\per\second}. We retrieved the frequencies of the expected lines at each redshift prior using the web-based \textsc{Splatalogue}\footnote{\url{https://www.cv.nrao.edu/php/splat/}} tool. This experiment resulted in no detections above $3\sigma$, other than the lines that were already identified (see Sect. \ref{sec:aca_data}). For the second experiment we proceed with a blind search of emission lines, via the automated \textsc{LineSeeker} software \cite{Gonzalez-Lopez2017ALMA-FFat1.1mm, Gonzalez-Lopez2019ASPECS-COLineScanAnd3mm}. Again, we found no detections within the SGASJ1226 and the \sbarc~Band 4 cubes. The software correctly identified two of the three lines in SGASJ0033 already detected by visual inspection: the bright $\co{4}{3}$ line associated with the arc (SGASJ0033-A) and the dimmer but broader emission feature at \SI{135.6}{\giga\hertz} toward the south east (SMG). \textsc{LineSeeker} did not detect the [\ion{C}{i}] line, however we consider it a reliable signal based on the strong priors on the frequency, position an width imposed by the $\co{4}{3}$ line. To confirm the detection, we extracted the [\ion{C}{i}] spectrum from the same spatial region as the CO line and convolve it with the best fit Gaussian profile of CO, resulting in a peak correlation $\sim\SI{25}{\kilo\meter\per\second}$ redward of CO's central velocity (see Fig. \ref{fig:ciline0033}).

\begin{figure*}[!htb]
    \centering
    \includegraphics[width=17cm]{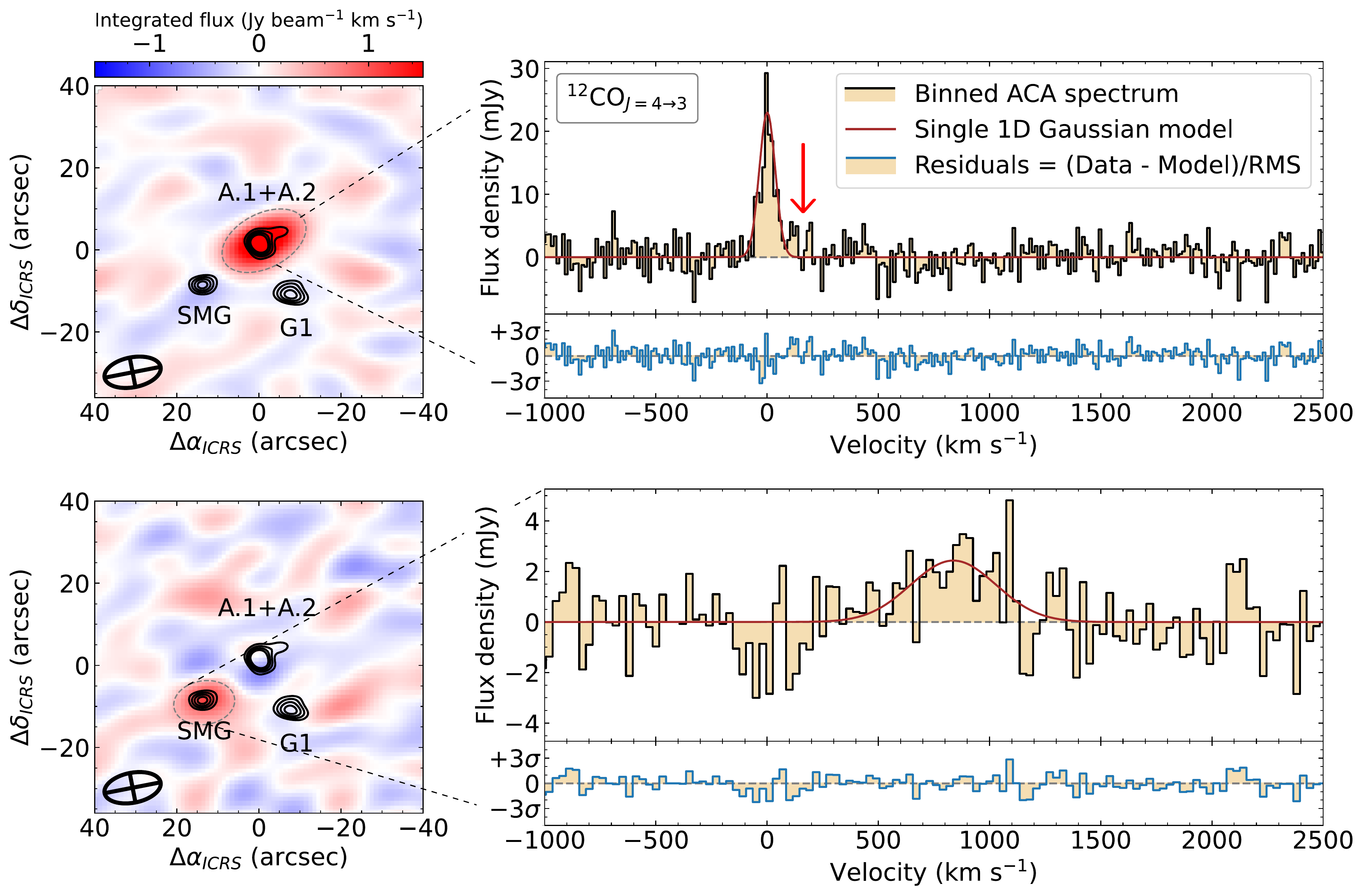}
    \caption{Zeroth moment maps and spectra of the emission lines detected in Band 4 toward SGASJ0033. The upper panels show the integrated brightness map of the $\co{4}{3}$ line (left) and its spectral profile (right) observed over images 1 and 2 of the lensed galaxy SGASJ0033-A at $z=2.39$. Black contours in both the upper left and lower left panels indicate the positive \submmw~continuum emission above $3\sigma$ already shown in Fig. \ref{fig:rb7_continua}. The black crossed ellipses indicate the size of the synthesized beam while the gray dashed ellipses show the apertures used for extracting the spectra shown in the corresponding right panel. The spectrum of $\co{4}{3}$ at A.1+A.2 (upper right panel) was resampled to a channel width of \SI{12.5}{\kilo\meter\per\second}; its best single Gaussian fit (see main text) is displayed in a red solid line. A tentative excess of emission at \SI{150}{\kilo\meter\per\second} is highlighted by a red vertical arrow. Normalized residuals from the difference between the observed spectrum and the model are shown in blue in the subplot below. The lower panels show the moment map (left) and spectrum (right) of the serendipitous emission line which we associate to the SMG. The channel width was resampled to \SI{30}{\kilo\meter\per\second} in order to maximize signal-to-noise while keeping five channels per FWHM. The spectral axis is in the same velocity frame as the spectrum of A.1+A.2 ($\nu_\text{rest}=\SI{135.996}{\giga\hertz}$). This choice highlights the proximity in velocity space of the two lines. }
    \label{fig:coline0033}
\end{figure*}

To extract line width and central velocity, we fit one dimensional Gaussian profiles to the CO spectra of SGASJ0033-A and SGASJ0033-SMG using non linear least squares minimization. Motivated by the marginal asymmetry of the line (see upper right panel of Fig. \ref{fig:coline0033}) and  the recent discovery of broad velocity components of nebular, rest-frame optical lines in this object \citep{Fischer2019outflowJ0033}, we tested for the presence of a similar feature in the observed CO profile. We fit single, double and triple Gaussian models, but only the single Gaussian achieved optimal corrected Akaike Information criterion \citep[AICc;][]{Cavanaugh1997CorrectedAkaikeCriterion} and Bayesian Information criterion \citep[BIC;][]{Wit2012StatsModelUncertainty} scores. These scores calculate a goodness-of-fit estimate (e.g.,~ $\chi^2$) but penalize overfitting by taking into account the number of degrees of freedom of the model. According to these criteria, double and triple Gaussian models are not as significant as the single Gaussian model. Thus, the results indicate no evidence of broad nor high velocity components within current data.

The total flux for each line was computed from the velocity integrated intensity maps (zeroth moment). Centered at the best fit central frequency, we used symmetric integration ranges spanning from $-0.85$ to $+0.85$ times the best-fit Gaussian FWHM. In this way we recover $\sim 99.5\%$ of the flux while keeping S/N$\gtrsim 3$ per channel. The resulting moment maps were then fit with a 2D Gaussian profile using CASA task \verb|imfit|, using the coordinates of the peak pixel as a prior on their position. The inferred fluxes with their respective uncertainties are listed in Table \ref{tab:lineflux}.  The results of the fits are consistent with the sources being unresolved by the ACA Band 4 beam. For the CO nondetections (SGASJ1226-A, SGASJ1226-B and \sbarc) we report $3\sigma$ point source upper limits assuming a fixed integration range width of \SI{200}{\kilo\meter\per\second} around the best spectroscopic redshift available.

\begin{figure}[!htb]
    \centering
    \includegraphics[width=\columnwidth]{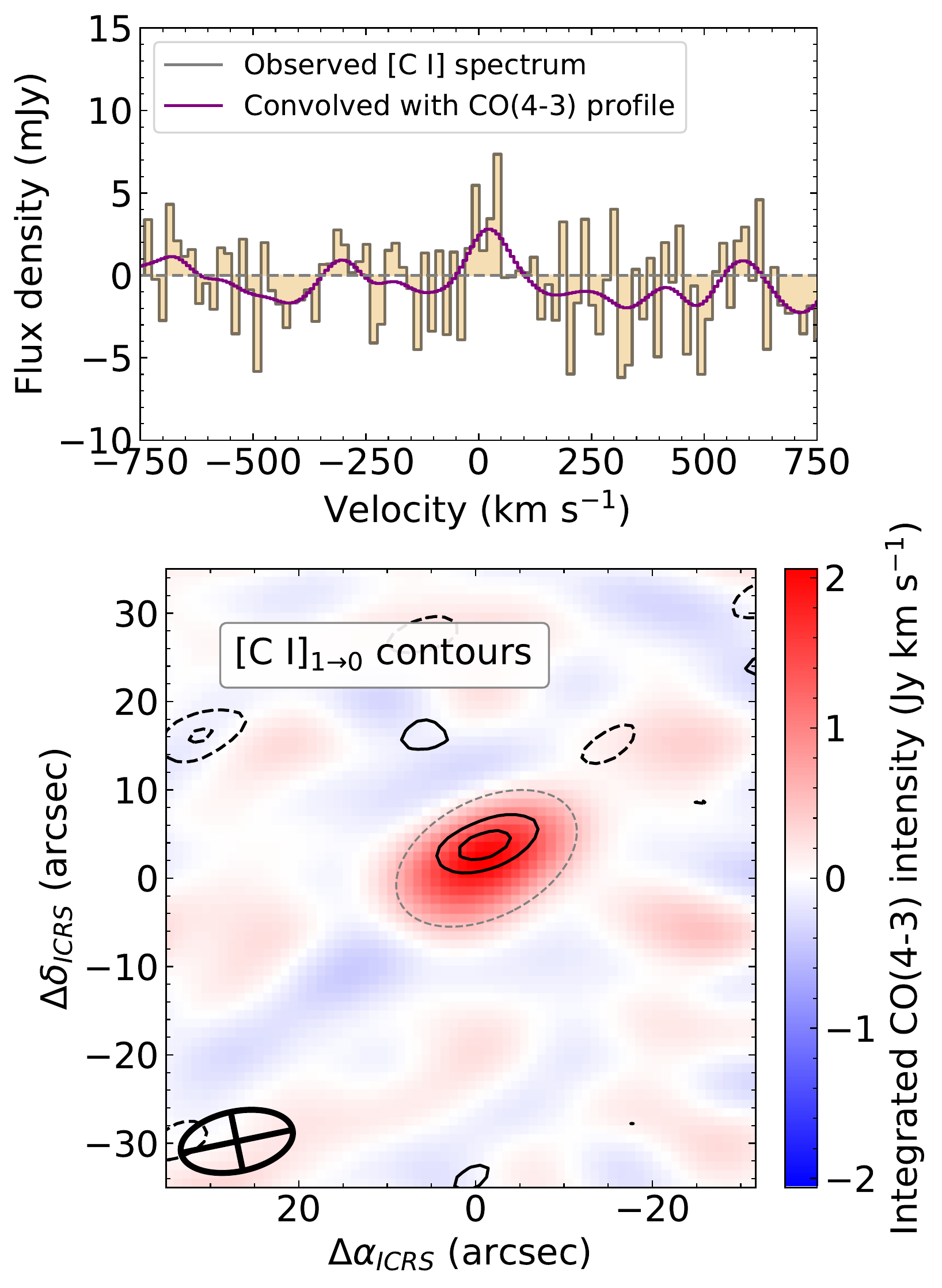}
    \caption{ACA Detection of [\ion{C}{i}]$_{1 \to 0}$ at $z=2.39$ toward the lensed galaxy SGASJ0033-A. \textit{Upper panel:} Continuum subtracted spectrum at \SI{145}{\giga\hertz} extracted from the A.1+A.2 aperture (dashed gray line in lower panel) with 2 channel binning (gray line). The overlaid purple curve is the result of the convolution between the observed spectrum and a Gaussian kernel with FWHM equal to the best fit value of the $\co{4}{3}$ line. The velocity axis is centered at the redshift of $\co{4}{3}$. \textit{Lower panel:} Integrated [\ion{C}{i}] emission contours (black solid/dahsed curves) at the  $\pm2,3\sigma$ levels on top of the $\co{4}{3}$ moment map. The crossed ellipse at the lower left corner indicates the size and orientation of the [\ion{C}{i}] beam.}
    \label{fig:ciline0033}
\end{figure}

\begin{table*}[!htb]
  \centering
  \caption{Emission line properties toward SGASJ0033}
  \begin{tabular}{cccccccc}
    \hline
    \hline
    Aperture & Central Frequency & Species & Redshift & FWHM & Bandwidth $\Delta v$ & $\mu \Delta v S_\nu$ & $\mu L^\prime_\text{line}$\\
             & \si{\giga\hertz} & & & \si{\kilo\meter\per\second} & \si{\kilo\meter\per\second} & \si{\jansky\kilo\meter\per\second} & $10^{10}$\,\si{\kelvin\kilo\meter\per\second\parsec\squared} \\
    \hline
    A.1+A.2 & $136.004 \pm 0.001$ & CO(4-3) & \num[separate-uncertainty=false]{2.3899+-0.000025} & $85 \pm 7$ & 164.2 & $1.96 \pm 0.15$ & $3.30 \pm 0.25$ \\
    A.1+A.2 & $145.176 \pm 0.004$ & [\ion{C}{i}] & \num[separate-uncertainty=false]{2.39010+-0.0009} & $75 \pm 26$ & 67.2 & $0.36 \pm 0.13$ & $0.53 \pm 0.19$ \\
    A.3 & $135.993 \pm 0.005$ & CO(4-3) & \num[separate-uncertainty=false]{2.39018+-0.00012} & $94 \pm 20$ & 135.4 & $0.66 \pm 0.22$ & $1.1 \pm 0.4$\\
    SMG & $135.643 \pm 0.017 $ & CO(4-3)\tablefootmark{a} & \num[separate-uncertainty=false]{2.3989+-0.0005} & $430 \pm 87$ & 412.7 & $1.42 \pm 0.33$ & $2.4 \pm 0.6$\\
    \hline
    \hline
  \end{tabular}
  \tablefoot{%
  \tablefoottext{a}{Tentative identification, assuming $z=2.4$}
  }
  \label{tab:lineflux}
\end{table*}

\subsection{Continuum photometry}
The flux of the \submmw~continuum detection was determined in a similar way to the CO line flux (see Sect \ref{sec:emission_lines}): we fit single 2D Gaussians to each source using \verb|imfit| on the cleaned continuum maps. In this case, the SGASJ0033 arc A.1 is resolved from the counter-image A.2 so we employed one Gaussian for each image. The (image plane) size of the best fit model for A.1 indicate that the arc is partially resolved along the major axis. Again, for the targets that were not detected, we quote the $3\sigma$ upper-limit for point sources based on the observed RMS and corrected by primary beam attenuation. We also measured the continuum flux on Band 4 ($\lambda_\text{obs}=\SI{2.1}{\milli\meter}$) data in a similar fashion. First we built dirty images after masking out the channels with detected and undetected line emission. Once again, SGASJ0033 is the only source with a reliable signal. The results are presented in Table \ref{tab:continuum_flux}. 

\subsection{SED fitting} \label{sec:sed}

\begin{figure}
    \centering
    \resizebox{\hsize}{!}{\includegraphics{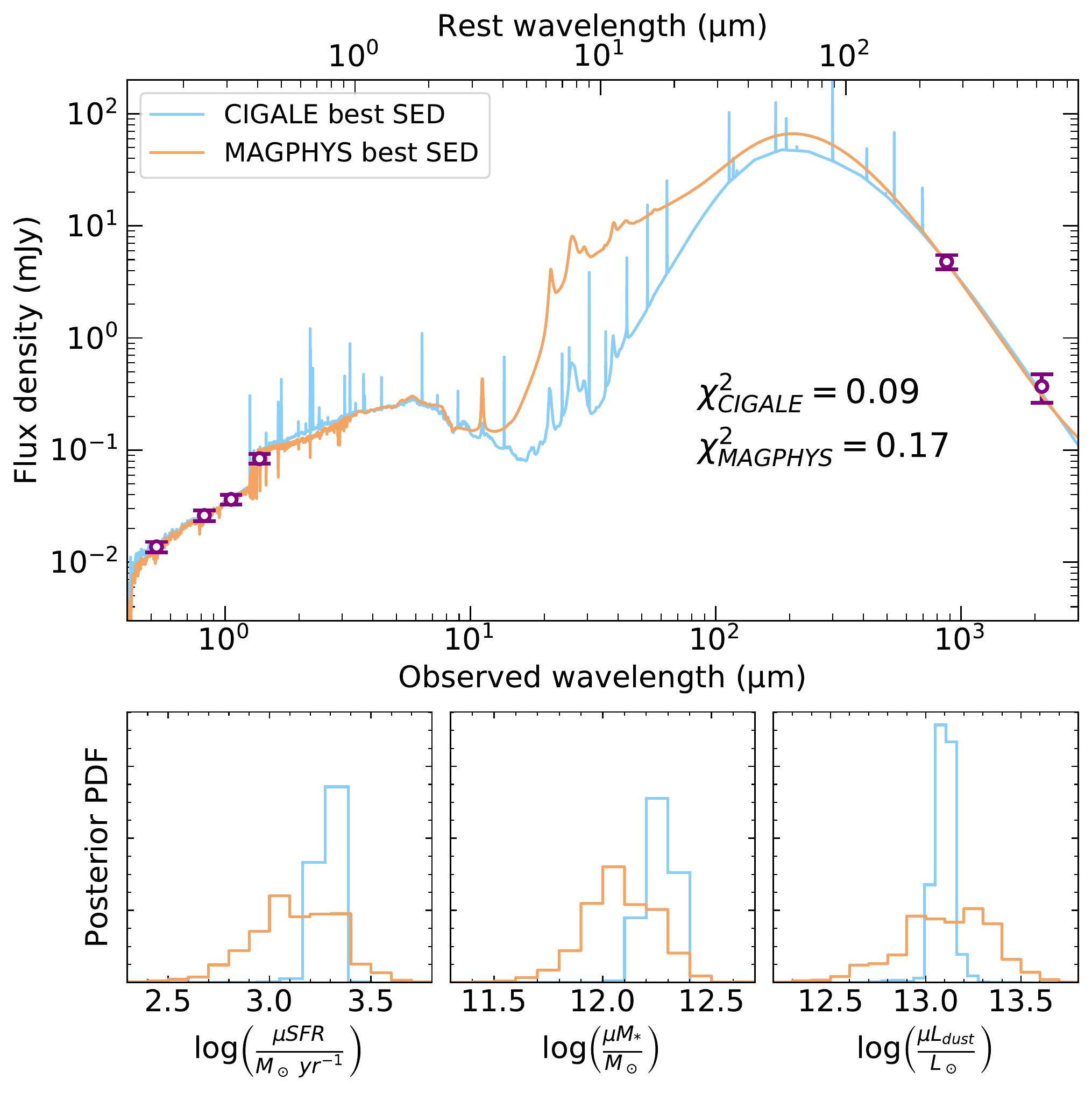}}
    \caption{Example SED fit for the lensed galaxy in SGASJ0033. Purple circles show the magnified image plane photometry from HST and ACA. The bottom panels show the posterior probability distributions for three model properties, comparing the \textsc{MAGPHYS} and CIGALE solutions. The lack of mid-infrared photometry leads to significant discrepancies between the two models between rest-frame \SI{2}{\micro\meter} and \SI{250}{\micro\meter}.}
    \label{fig:sed_j0033}
\end{figure}
We combined image plane optical HST photometry (four bands), ACA submillimeter continuum photometry (two bands) and IRAC near infrared photometry (two bands) when available to fit templates of spectral energy distributions (SEDs) and derive estimates of stellar mass and star formation rate. For this purpose, we used the high-$z$ version of the Multi-wavelength Analysis of Galaxy Physical Properties package \citep[\textsc{Magphys};][]{daCunha2015HighZMagphysALESS}. \textsc{Magphys} uses precomputed grids of stellar emission \citep{Bruzual&Charlot2003SSP} and dust emission models \citep{daCunha2008MagphysPaper} to evaluate the likelihood that the observed multi-band photometry comes from each combination of stars plus dust models. The results are given as samples of the posterior probability distribution of the parameters of the models. The code operates under the principle of energy balance, which states that all the energy of the UV radiation field absorbed by the dust in the interstellar medium is re-emitted in the far infrared  as a thermal continuum. When both the UV and FIR fluxes are observed, and a given attenuation law is assumed \citep[in the case of \textsc{Magphys},][]{Charlot&Fall2000DustAttenuation}, this principle provides a useful constraint that can break the degeneracies associated with galaxy reddening. Still, with no data in the mid infrared and a low total number of bands observed, SED fitting suffers from large uncertainties (e.g., from the unconstrained contribution of AGN). In order to address possible systematics we also ran  \textsc{CIGALE} \citep{Boquien2019CIGALEpaper}, another SED fitting code, and then compare with our \textsc{Magphys} results. \textsc{CIGALE} allows more flexibility in the modeling choices, like the shape of the star formation history (SFH), the attenuation law or the inclusion of other components like radio synchrotron emission. For this experiment, we used the same stellar templates and attenuation law as \textsc{Magphys}, but assumed a delayed exponential SFH and a nebular emission component \citep[based on templates by][]{Inoue2011NebularTemplates}. We also used the simpler dust emission model proposed by \citet{Dale2014DustTemplate}, which is a grid of templates that depend on two parameters: $\alpha_\text{SF}$, which governs the strength of the dust heating intensity, and $f_\text{AGN}$, the AGN fraction. Here we have chosen to fix $f_\text{AGN}=0$ and vary $\alpha_\text{SF}$ between 0 and 4. We found that both \textsc{CIGALE} and \textsc{Magphys} offer reasonable fits to the observed bands, but the former produces on average $\sim 0.15$ dex  and $\sim 0.3$ dex higher estimates of SFR and $M_\text{stars}$ respectively than \textsc{Magphys}. These differences are consistent with the results of a comparison between \textsc{Magphys} masses and EAGLE/SKIRT simulated galaxies \citep{Dudzeviciute2020AS2UDS}.  However, the Bayesian posterior probability distributions of these parameters produced by either codes do overlap within their 68\% confidence intervals (see Figure \ref{fig:sed_j0033} for an example), so we do not consider the differences to be statistically significant. In what follows, we report \textsc{Magphys} derived values corrected by magnification (see Table \ref{tab:sed_results}).

\begin{table*}[!htb]
    \centering
    \caption{\label{tab:sed_results} Physical properties derived from SED fitting and dust masses from far-infrared photometry}
    \input{sed_results}
    \tablefoot{%
    The second column shows the fiducial magnification factor applied to bring values into the source plane. Below the horizontal rule, we add (for completeness) the results for the foreground galaxies detected in Band 7 except for the SMG which has a single detection in the near infrared and lacks a spectroscopic redshift.  The last two columns, separated by a vertical line, contain the dust masses inferred from the continuum measurements in bands 7 and 4. 
    
    \tablefoottext{a}{Fluxes were measured on partial lensed images.}
    \tablefoottext{b}{Fitted at fixed $z=0.77$ \citep{Tejos2021ArctomoJ1226Kinematic}}
    \tablefoottext{c}{Fitted at fixed $z=1.17$ (priv. comm. with C. Ledoux).}
    }
\end{table*}

\subsection{Dust mass} \label{sec:dust_mass}
The MAGPHYS fits to the far infrared SED deliver posterior probability distributions of the total dust mass, but since that part of the SED is sparsely sampled (single band detection or upper limits) the uncertainties are large. An alternative way to estimate the dust mass directly from the single band sub-millimetric flux is by using the framework proposed by \citet{Dunne2000SLUGS}. This method is based on the assumption that optically thin dust traces the cold ISM, and therefore the observed continuum flux is proportional to the dust mass \citep[e.g.,][]{Eales2012DustEmissionAsMassProbe, Scoville2014DustEmissionAsGasProbeEvolution,Scoville2016MainSequenceAndGasMass}. Following \citet{Scoville2016MainSequenceAndGasMass}, we set the global mass-weighted dust temperature to $T_d=\SI{25}{\kelvin}$ and the dust emissivity index to $\beta=1.8$. Then, the dust mass is computed as,
\begin{equation}
    \text{M}_\text{dust} = \frac{S_{\nu_\text{obs}} D_L^2 (1+z)^{-(3+\beta)}}{\kappa(\nu_\text{ref}) B_\nu(\nu_\text{ref}, T_d)} \left(\frac{\nu_\text{ref}}{\nu_\text{obs}}\right)^{2+\beta}\left(\frac{\Gamma_{\text{RJ(ref, 0)}}}{\Gamma_\text{RJ}}\right),
\end{equation}
where $\nu_\text{obs}$ is the observed frequency, $S_{\nu_\text{obs}}$ the measured flux and $D_L$ the luminosity distance. The quantity $\kappa(\nu_\text{ref})$ is the dust mass absorption coefficient at rest frame $\nu_\text{ref}$; here we adopted $\kappa(\SI{450}{\micro\meter})=\SI{1.3}{\centi\meter\squared\per\gram}$ \citep{Li&Draine2001DustISM}. $B_\nu$ is the Planck function and $\Gamma_\text{RJ}\left(\nu,\, z,\, T_d\right)$ is a factor that accounts for the deviation of the Planck function from the Rayleigh-Jeans (RJ) law. The intrinsic dust masses (or upper limits) obtained with this method are presented along the SED results in Table \ref{tab:sed_results}.

\begin{figure}
    \centering
    \includegraphics[width=\columnwidth]{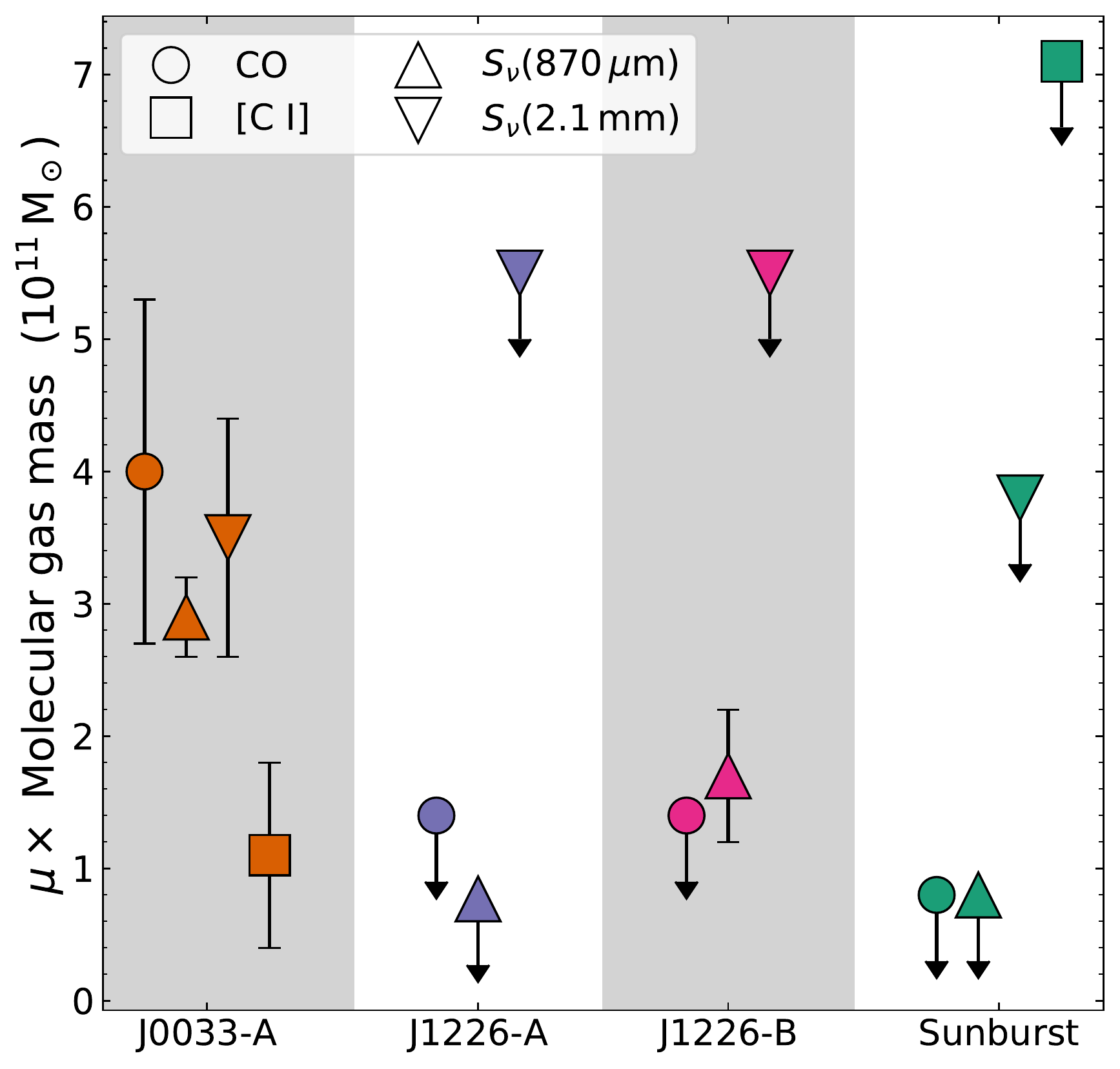}
    \caption{Measurements and upper estimates on image plane molecular gas mass according to different tracers. The gray background strips serve as a visual aid to separate the different sources. The molecular gas mass measurements of SGASJ0033 based on CO and dust continuum are consistent with each other, while the [\ion{C}{i}] based estimate favors a lower value.}
    \label{fig:tracer_comparison}
\end{figure}

The validity of this method relies on the restriction $\lambda_\text{rest}>\SI{250}{\micro\meter}$, which ensures that one is probing the Rayleigh-Jeans tail and dust is optically thin \citep{Scoville2016MainSequenceAndGasMass}. Here, the condition is met for SGASJ0033 and the \sbarc, but not for the SGASJ1226 galaxies since \submmw~at $z=2.92$ becomes \SI{222}{\micro\meter}. However, low metallicity galaxies --like SGASJ1226-A and B-- typically have higher dust temperatures than their solar metallicity equivalents at fixed redshift \citep{RemyRuyer2013DwarfGalaxySurveyPhotometry, Saintonge2013MolGasLensedGalaxies}. In such cases, the SED peak and Rayleigh-Jeans tail are shifted toward shorter wavelengths, hence relaxing the $\lambda_\text{rest}$ restriction. Also, the calibration of \citet{Scoville2016MainSequenceAndGasMass} did include \submmw~flux measurements of SMGs at $z\gtrsim 2.5$, since they were found to follow the same $L^\prime_\text{CO}-L_\nu(\SI{850}{\micro\meter})$ trend as local spirals and ULIRGs. For these two reasons, we believe that $\lambda_\text{obs}=\submmw$ remains a good tracer of dust mass for the two lensed galaxies in the SGASJ1226 system, and the method outlined above still applies.  In any case, we repeated the measurement with the Band 4 continuum  ($\lambda_\text{obs}=\SI{2.1}{\milli\meter}$), obtaining a very similar value for SGASJ0033 as with \submmw~($\mu M_\text{dust}^{\submmw}=\SI{2.5+-0.3e9}{\msun}$ vs $\mu M_\text{dust}^{\SI{2.1}{\milli\meter}}=\SI{3.0+-0.8e9}{\msun}$), though with higher uncertainty. Despite tracing a rest-wavelength that is well within the RJ regime and closer to the reference wavelength of \SI{850}{\micro\meter}, the \SI{2.1}{\milli\meter} continuum is expected to be much fainter than \submmw~at these redshifts, hence the upper limits obtained with Band 4 measurements are less restrictive (see Fig. \ref{fig:tracer_comparison}).

\subsection{Molecular gas mass} \label{sec:gas_mass}

A major goal of this paper is to constrain the molecular gas content of the source galaxies producing the giant arcs. To estimate the global molecular gas mass, we employ at least one of the following methods:
\begin{enumerate}[(i)]
    \item \textit{Based on \co{J}{J-1} luminosity}: The observed CO line flux can be converted to the ground transition \co{1}{0} luminosity, which is the traditional and best calibrated H\textsubscript{2} tracer. To correct for the excitation level, here we adopt the median ratios $r_{J1}=L^\prime_{\text{CO}(J\to J -1)} / L^\prime_{\text{CO}(1 \to 0)}$ from \citet{Kirkpatrick2019COEmissionAGN} calibrated in a sample of matched redshift and intrinsic infrared luminosity. Once $L^\prime_{\text{CO}(1\to 0)}$ is computed, we apply the \citet{Genzel2015DustAndCOscalingRelations} recipe for estimating the CO-to-H\textsubscript{2} conversion factor (hereafter $\alpha_\text{CO}$) based on gas-phase metallicity. Without a homogeneous metallicity indicator for every target, we adopt the stellar metallicities reported by \citet{Chisholm2019MegaSauraMetallicity} as a proxy. A discussion on the validity of this choice can be found in Appendix \ref{sec:metallicity}.
    \item \textit{Based on M\textsubscript{dust}:} Giant molecular clouds in the ISM of star-forming galaxies contain cold dust that can be detected in long-wavelength thermal continuum emission \citep{Eales2012DustEmissionAsMassProbe, Magdis2012EvolvingISMatRedshiftTwo}. With a measurement of the dust mass available (see Sect. \ref{sec:dust_mass}), one can use the gas-to-dust mass ratio ($\delta_\text{GDR} = M_\text{mol}/M_\text{dust}$) to infer the total molecular mass \citep{Scoville2016MainSequenceAndGasMass}. Here we adopt a scaling with gas-phase metallicity as $\delta_\text{GDR}\propto Z^\gamma$, normalized at  $\delta_\text{GDR}(Z_\odot)=100$ \citep{Draine2007Dust} and a power-law index $\gamma=-0.85$ taken from \citet{Tacconi2018PHIBSS2MainSequence}. This calibration should be valid for solar to slightly subsolar metallicities \citep{Leroy2011COtoH2, Remy-Ruyer2014GasToDustLocalGalaxies}, but will be revisited in Sect. \ref{sec:co_deficit}. As presented in Sect. \ref{sec:dust_mass}, we derive dust mass from both Band 7 (\submmw) and Band 4 (\SI{2.1}{\milli\meter}) continuum.
    
    \item \textit{Based on \neutralc~luminosity:} The fine structure lines of atomic carbon have been proposed as reliable tracers of cold gas \citep[e.g.,][]{PapadopoulosAndGreve2004CIasGasTracer, Danielson2011StarformingGalaxyAtRedshift2, Popping2017CIandCOatZ2}. In order to convert \neutralc~luminosity to H\textsubscript{2} mass one first has to compute the total atomic carbon mass. Since both $\element[][3]{P_2}\to\element[][3]{P_1}$ and $\element[][3]{P_1}\to\element[][3]{P_0}$ lines are optically thin and have low critical densities, the mass only depends on the luminosity and excitation temperature $T_\text{ex}$ \citep{Weiss2003GasAndDustCloverleaf, Weiss2005AtomicCarbon}. The excitation temperature can be estimated from the luminosity ratio of both transitions, but since our data only covers the $\element[][3]{P_1}\to\element[][3]{P_0}$ transition, we set $T_\text{ex}=\SI{30}{\kelvin}$ as typically assumed in the literature \citep[e.g.,][]{Bothwell2017AtomicCarbonSPT, Popping2017CIandCOatZ2, Valentino2018SurveyOfAtomicCarbon, Brisbin2019NeutralCarbonAndCOMassiveSFG, Boogaard2020ASPECS-COExcitation, Boogaard2021ColdGasLowMassSFGs}. At $T_\text{ex}=\SI{30}{\kelvin}$, a 20\% temperature variation produces  a $\lesssim 1\%$ variation in M\textsubscript{[CI]} \citep{Weiss2005AtomicCarbon, Boogaard2020ASPECS-COExcitation}. Once the atomic carbon mass has been computed (i.e., with Eq. (1) of \citealt{Weiss2005AtomicCarbon}), we use the metallicity dependent prescription of \citet{Heintz&Watson2020CItoH2} for the carbon abundance to convert to molecular gas mass. Unlike method (ii) whose $\delta_\text{GDR}(Z)$ scaling is calibrated on CO-derived gas masses, the   \citet{Heintz&Watson2020CItoH2} relation was obtained directly from the [\ion{C}{i}]/H\textsubscript{2} column density ratio observed in a sample of quasar and gamma ray burst absorbers at high-$z$, and hence is independent of the $\alpha_\text{CO}$ factor. For this reason, the [\ion{C}{i}]-based M\textsubscript{mol} provides useful consistency checks on the previous two methods.
\end{enumerate}

Since SGASJ0033-A has detections of CO, [\ion{C}{i}] and dust continuum, we can use it as a benchmark for the three methods listed above. Following method (i) we adopted $r_{41}=0.37\pm0.12$ \citep[from][]{Kirkpatrick2019COEmissionAGN} and $\alpha_\text{CO}=\SI{4.5+-0.1}{\msun}\,(\si{\kelvin\kilo\meter\per\second\parsec\squared})^{-1}$  based in our fiducial metallicity $Z/Z_\odot=\num{0.84+-0.02}$ (Appendix \ref{sec:metallicity}). We then  converted the observed image plane CO flux into $\mu M_\text{mol, CO}^{A.1+A.2}=\SI{4.0+-1.3e11}{\msun}$, where $A.1+A.2$ indicates that the value considers the blended flux of the two lensed images. With $\delta_\text{GDR}(Z)=117$ and $X_\text{[CI]}(Z)=\num{1.3+-0.8e-5}$ (for reference, the typical value assumed for high-$z$ studies is \num{3e-5}, \citealt{Valentino2018SurveyOfAtomicCarbon}) results from methods (ii) and (iii) are $\mu M_\text{mol, dust}^{A.1+A.2} = \SI{2.9+-0.3e11}{\msun}$ and $\mu M_\text{mol, [CI]}^{A.1+A.2} = \SI{1.1+-0.7e11}{\msun}$ respectively. While the values obtained with methods (i) and (ii) are in reasonable agreement for both tracers of M\textsubscript{dust}, the (iii) method yields an estimate that is $\sim 2\sigma$ lower. This can be a result of the low significance of the [\ion{C}{i}] detection or a systematic effect driven by the uncertain carbon abundance factor. In what follows, we adopt the CO estimate as the fiducial value of M\textsubscript{mol} for SGASJ0033-A.

\section{Discussion} \label{sec:discussion}
\subsection{Lensed galaxies in context} \label{sec:scaling_relations}
In order to locate our derived ISM properties of lensed galaxies within the context of high-$z$ scaling relations, we compare our results with a pair of reference samples. The main reference sample is the latest release of the IRAM Plateau de Bure High-z Blue Sequence Survey \citep[PHIBSS 1/2;][]{Tacconi2013PHIBSS1,Tacconi2018PHIBSS2MainSequence} containing 1444 molecular gas measurements at $0\leq z \leq 4$ and covering a stellar mass range from $\log\left(\text{M}_\text{stars}/\text{M}_\odot\right)=9$ to 11.9. The sample includes both individual galaxies and stacked measurements from surveys with high detection rates. The sample was selected to represent the overall SFG population in a wide range of basic galaxy parameters. The bulk of the sample are galaxies that belong to the star-forming main sequence, with a modest contribution from star-bursting outliers \citep{Tacconi2018PHIBSS2MainSequence}. From this parent sample we selected two subsets: firstly, a high-$z$ sample defined as all the sources in the  PHIBSS 1/2 catalog with spectroscopic redshift greater than 2, excluding lensed galaxies. This subset contains 138 objects with stellar masses between \SI{e9.8}{\msun} and \SI{e11.8}{\msun} and a median redshift of 2.3. The molecular gas masses were derived from CO luminosity for 59 sources and from dust continuum for the other 79 sources. Secondly, the entire xCOLD GASS sample \citep{Saintonge2011xCOLDGASSp1} as appears in PHIBSS 1/2. Contains 306 galaxies at $z\sim 0$ with CO measurements and stellar masses ranging from \SIrange{e9}{e11.3}{\msun}.

We also compiled a sample of molecular gas measurements in UV-bright strongly-lensed galaxies from \citetalias{Saintonge2013MolGasLensedGalaxies} and \citetalias{DessaugesZavadsky2015MolGasLensedSFGs}. These two studies provide global properties and molecular gas masses for several giant arcs, most of them selected from optical surveys. In both studies, a complete sampling of the infrared SED with \textit{Spitzer} and \textit{Herschel} is combined with dedicated PdBI or IRAM 30m CO line observations to constrain the parameters of the cold ISM of these galaxies. After we remove duplicate objects between the two samples and exclude the SGASJ1226-A arc from the \citetalias{Saintonge2013MolGasLensedGalaxies} table, this sample comprises 17 sources at mean redshift of $z\approx 2.1$ and a de-lensed stellar mass range of \SIrange{e9.3}{e11.5}{\msun}. 

Each sample uses their own set of calibrations and tracers for the molecular gas. On one hand, the PHIBSS catalog uses the \citetalias{Genzel2015DustAndCOscalingRelations} metallicity-dependent recipe for estimating $\alpha_\text{CO}$, while \citetalias{Saintonge2013MolGasLensedGalaxies} used the \citealt{Genzel2012MetallicityCOtoH2} \citepalias{Genzel2012MetallicityCOtoH2} recipe. On the other hand, \citetalias{DessaugesZavadsky2015MolGasLensedSFGs} assumed a constant Galactic value of $\alpha_\text{CO} = 4.6\,\si{\msun}(\si{\kelvin\kilo\meter\per\second\parsec\squared})^{-1}$. Here, we attempted to standardize the reference samples to a common scheme for unbiased comparison with our sample. We updated the $M_\text{H2}$ values in Table 6 of \citetalias{Saintonge2013MolGasLensedGalaxies} to the \citetalias{Genzel2015DustAndCOscalingRelations} calibration based on the metallicities provided in their Table 5. With this calibration, the molecular gas mass estimations become 0.2 dex lower on average. For the \citetalias{DessaugesZavadsky2015MolGasLensedSFGs} sample we retrieved metallicities from \citetalias{Saintonge2013MolGasLensedGalaxies} for the duplicate sources cB58 and the \textit{Cosmic Eye}. For the rest of their sample, we computed metallicities from the MZR as appears in \citetalias{Genzel2012MetallicityCOtoH2} and then converted to the \citet{Denicolo2002MetallicityCalibration} scale for consistency with \citetalias{Saintonge2013MolGasLensedGalaxies}. We finally applied the \citetalias{Genzel2015DustAndCOscalingRelations} recipe to get $\alpha_\text{CO}$ and updated the reported values of $M_\text{mol}$ accordingly. This resulted in $\sim 0.1$ dex higher gas mass estimations relative to their published values.

 We also checked that the \textsc{Starburst99} stellar metallicities from \citet{Chisholm2019MegaSauraMetallicity} in our sample are consistent with the expectations from the mass metallicity relation (MZR; see Sect. \ref{sec:intro}): we fed the SED-derived stellar masses into the MZR recipe used by \citetalias{Genzel2012MetallicityCOtoH2} and converted them to the \citet{Denicolo2002MetallicityCalibration} scale. The resulting oxygen abundances are all within 0.2 dex of the \citet{Chisholm2019MegaSauraMetallicity} value. Thus, even if the intrinsic scatter of the MZR is lower than 0.2 dex at these resdhifts, the deviations we observe are fully consistent with our $M_\text{stars}$ uncertainty.

\begin{figure}[!htb]
    \centering
    \resizebox{\hsize}{!}{\includegraphics{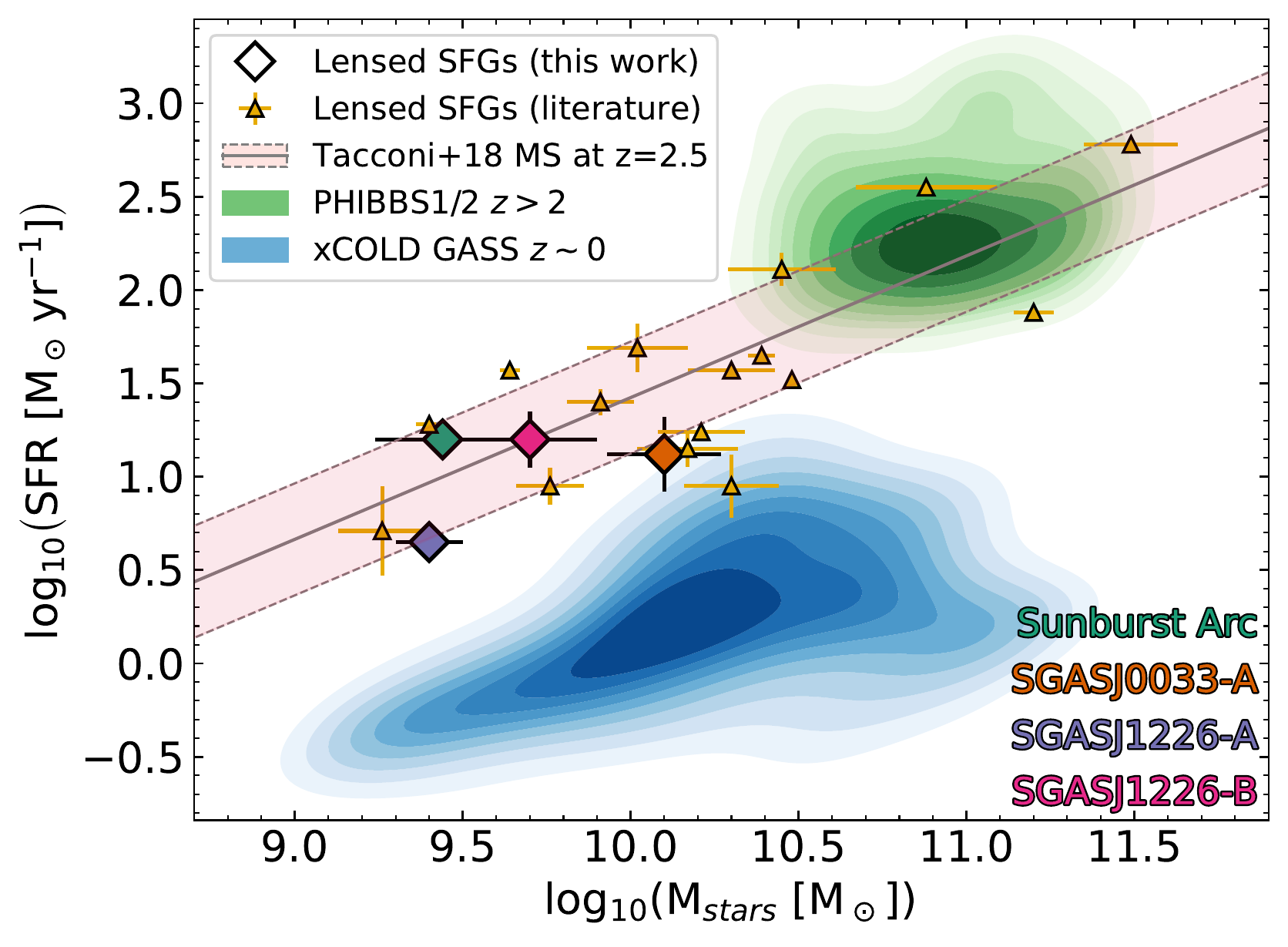}}
    \caption{Star formation rate versus stellar mass diagram for the lensed sources in this work, compared to literature samples. The PHIBSS1/2 subsamples are displayed as filled contours of Gaussian kernel density estimates, starting at the 20\%  peak density level. The solid line indicates the empirical location of the star-forming main sequence at $z=2.5$ derived from PHIBSS1/2 data by \citet{Tacconi2018PHIBSS2MainSequence}. The dashed lines and pale red filling indicate the typical $\pm0.3$ dex scatter in this relation. Orange-filled triangles represent lensed galaxies from \citep{Saintonge2013MolGasLensedGalaxies} and \citep{DessaugesZavadsky2015MolGasLensedSFGs}.} 
    \label{fig:ms}
\end{figure}

Figure \ref{fig:ms} shows the ranges of stellar mass and star formation rate that become accessible with the help of strong gravitational lensing. Compared to the unlensed high-$z$ PHIBSS1/2 sample, the lensed galaxies occupy a much wider range in both SFR and $M_\text{stars}$, while still being located near (within $1\sigma$ errors) the empirical MS proposed in \citetalias{Tacconi2018PHIBSS2MainSequence}. In particular, the four sources in our sample have stellar masses well within the range of the $z\sim 0$ SFGs represented by the xCOLD GASS sample, a regime not yet explored by unlensed surveys at this redshift.

The ratio between the total molecular gas mass and the total stellar mass (hereafter gas fraction, $M_\text{mol} / M_\text{stars}$) is a crucial parameter to characterize the galaxy-integrated ISM properties, since it relates the amount of gas available to produce stars in relatively short timescales with the accumulated stellar mass buildup. There is a growing consensus that high-redshift galaxies have larger gas fractions than local galaxies for a given stellar mass \citep[e.g.,][]{Scoville2017EvolutionISM, Tacconi2018PHIBSS2MainSequence, Liu2019A3COSMOScoldGasEvolution}, a fact that might be explained by increased rates of gas accretion from the cosmic web at earlier times \citep[e.g.,][]{Dekel2009ColdModeAccretion, Walter2020EvolutionOfBaryons}. However, the current surveys of molecular gas at high redshift are not sensitive to lower mass ($M_\text{stars} \lesssim \SI{e10}{\msun}$) systems \citep[see][for a review]{Hodge&daCunha2020Review}, so it remains unclear whether galaxies in this regime follow the same trends \citep[e.g.,][]{Coogan2019SuppressedCOinLowMetGals, Boogaard2021ColdGasLowMassSFGs}.

\begin{figure*}[!hbt]
    \centering
    \includegraphics[width=17cm]{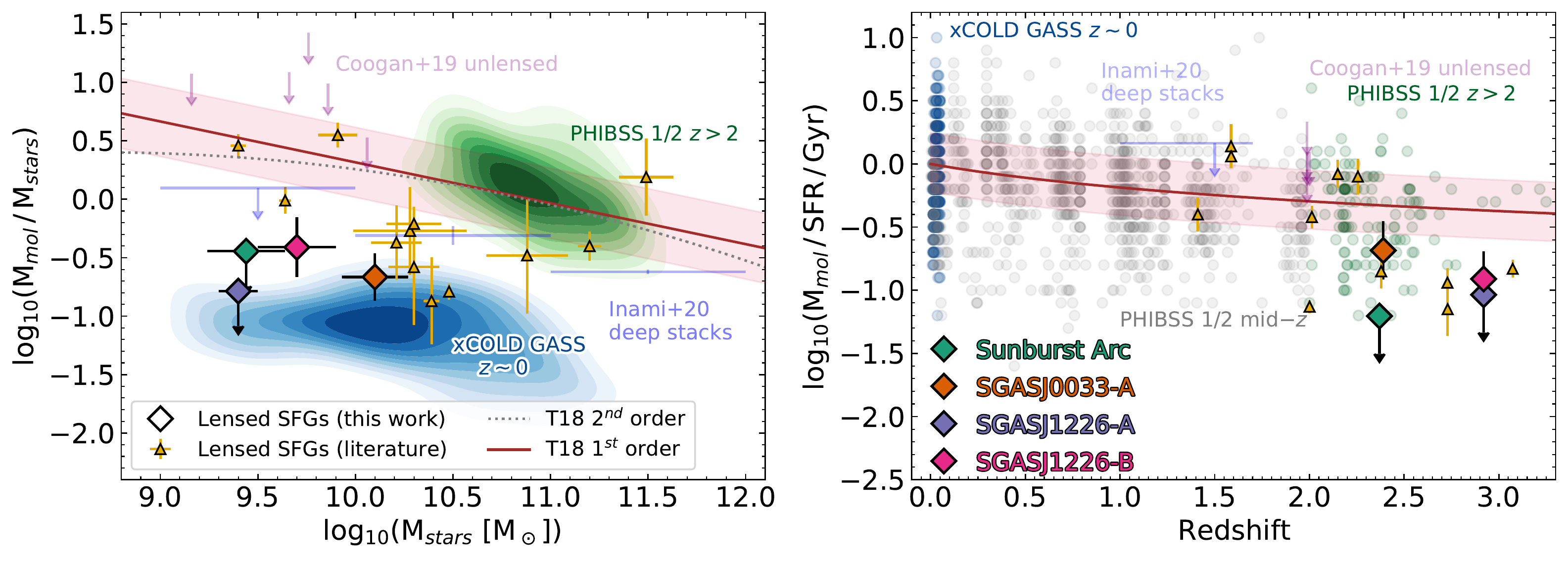}
    \caption{Molecular gas fraction and depletion timescale (M\textsubscript{mol}/SFR) in the context of high and low redshift galaxies. We only show our values based on the \submmw~band continuum as a tracer of M\textsubscript{mol}, since it has the highest sensitivity and shows remarkable agreement with the CO-based values (see Fig.~\ref{fig:tracer_comparison}). The only exception is SGASJ0033-A for which we plot the CO-based value. \textit{Left:} The filled green (blue) contours show the locus of the Gaussian kernel density estimator for the PHIBSS 1/2 $z>2$  (xCOLD GASS) sample.
    Squares indicate the location of the lensed galaxies presented in this paper, identified by color in the right panel.  We also include the joint sample from \citetalias{DessaugesZavadsky2015MolGasLensedSFGs} and \citetalias{Saintonge2013MolGasLensedGalaxies} as orange filled triangles.
    The brown solid and gray dotted lines show the \citet{Tacconi2018PHIBSS2MainSequence} scaling relation at $z=2.5$ using the first and second order fits respectively.
    Pink downward arows at the top left are the upper limits from the CO nondetections presented in \citep{Coogan2019SuppressedCOinLowMetGals}, while the pale blue errorbars are the result of deep CO stacking in the \textit{Hubble} Ultra Deep Field with $1.0 < z_\text{spec} < 1.7$ in bins of one dex in M\textsubscript{stars} \citep[][the bin with $8< \log_{10}\left(\text{M}_\text{stars}/\text{M}_\odot\right) < 9$ is not shown]{Inami20202StacksAspecsCO}.
    \textit{Right:} Gas depletion time versus redshift. The solid brown line is the  \citetalias{Tacconi2018PHIBSS2MainSequence} scaling relation for galaxies with $M_\text{stars}=\SI{e9.7}{\msun}$, the median mass of our sample. PHIBSS 1/2 data is displayed as shaded circles instead of contours, but the darkness of each circle is a proxy of the number density. Again, the lensed sources studied here fall below the expected relation. The stacked upper limit from \citet{Inami20202StacksAspecsCO} is also shown, and
    corresponds to the M\textsubscript{stars} bin between \SI{e9}{\msun} and \SI{e10}{\msun}}
    \label{fig:moneyplot}
\end{figure*}

In Fig. \ref{fig:moneyplot}, we compare the gas fractions of our sample with the literature samples mentioned above. We observe that the $z\sim2$ lensed galaxies do, in fact, have larger gas fractions than the local xCOLD GASS sample, but these values are lower than expected for their redshift and mass range based on the extrapolation of the \citetalias{Tacconi2018PHIBSS2MainSequence} relation. Our detections and upper limits reach values that are below the sensitivities of individual ALMA observations of low mass galaxies at $z\approx 2$ \citep[pink arrows;][]{Coogan2019SuppressedCOinLowMetGals} and deep stacked measurements from the ASPECS survey \citep[pale blue errorbars;][using the values of the "on-MS" row of their Table 5]{Inami20202StacksAspecsCO}.

In the left panel we see that gas fraction is anticorrelated with stellar mass for the high-$z$ PHIBSS 1/2 galaxies, that is, more massive galaxies have relatively smaller gas reservoirs. In the stellar mass interval between \SI{e10.3}{\msun} and \SI{e11.5}{\msun}, both the high-$z$ and the local sample follow similar trends, but at lower masses, the local sample reveals a flattening of the relation.  According to \citetalias{Tacconi2018PHIBSS2MainSequence}, this nonlinear behavior can be interpreted as a manifestation of the "mass-quenching" feature of the main sequence \citep[suppression of SFR at high M\textsubscript{stars}, e.g.,][]{Whitaker2014MainSequence, Schreiber2015MainSequence}, though it has also been suggested that this is just a selection effect due to mass incompleteness \citep{Liu2019A3COSMOScoldGasEvolution}. Here we show that the lensed galaxies distribute at $\log(M_\text{mol}/M_\text{stars})\lesssim - 0.5$, with no clear indication of M\textsubscript{stars} dependence. Although the scatter is large and the number of sources small, the lensed galaxies have, on average, gas fractions one order of magnitude below the value one would obtain by extrapolating the linear \citetalias{Tacconi2018PHIBSS2MainSequence} relation beyond the PHIBSS 1/2 mass range at $z>2$. In detail, the offsets from the linear \citetalias{Tacconi2018PHIBSS2MainSequence} (red line in Fig. \ref{fig:moneyplot}) relation are 0.7, 0.8, $>1.2$ and $>1.1$ dex for SGASJ0033-A, SGASJ1226-B, SGASJ1226-A and the \sbarc, respectively. An alternative second-order relation is also given in \citetalias{Tacconi2018PHIBSS2MainSequence} (dotted gray line in Fig. \ref{fig:moneyplot}), which accounts for the low mass flattening, but even with this correction the lensed galaxies show a lower gas fraction than predicted.

The right panel of Fig. \ref{fig:moneyplot}, shows a comparison of the gas depletion timescale ($M_\text{mol}/\text{SFR}$) as a function of redshift. This plot also includes the intermediate-redshift sources we had excluded from the PHIBSS 1/2 subsample. While the scatter of points with respect to the \citetalias{Tacconi2018PHIBSS2MainSequence} scaling relation is large ($0.4$ dex), the lensed galaxies from this work present gas depletion timescales that are shorter than expected based on the \citetalias{Tacconi2018PHIBSS2MainSequence} relation by 0.6, 0.6, $>0.8$ and $>0.8$ dex for SGASJ0033-A, SGASJ1226-B, SGASJ1226-A and the \sbarc, respectively. SGASJ0033-A is in  agreement with other (albeit more massive) sources at the same redshift, but the tension is higher for the nondetected \sbarc~and SGASJ1226-A, whose gas will deplete in less than \SI{100}{\mega\year} at the current SFR. 
Taken at face value, these results suggest that the lensed galaxies track a gas-deficient population that contrasts with the more massive galaxy population at high-redshift. In the following, we discuss on possible biases affecting the measurements.

\subsection{The molecular gas deficit cannot be explained by systematics in M\textsubscript{stars} or SFR} \label{sec:systematics}
Before we can ascribe the observed deficit in the gas fraction and depletion time to the properties of the sampled galaxy population, we explore potential systematic effects on the measured quantities that can explain the discrepancy. For this, we temporarily adopt the hypothesis that our sample should strictly follow the \citetalias{Tacconi2018PHIBSS2MainSequence} scaling relations. Under this hypothesis, the low gas fraction and depletion time can be driven, for example, by a systematic underestimation of the molecular gas mass, or conversely, by an overestimation of the stellar mass and SFR. In this section we focus on the latter effect, while a discussion on a possible underestimation of molecular gas mass is postponed to section \ref{sec:co_deficit}.

A shift on the measured stellar mass (keeping everything else constant) would need to be on the order of 0.7-1.2 dex to reconcile the observed gas fraction with \citetalias{Tacconi2018PHIBSS2MainSequence}. Some degree of offset in this direction could be achieved, for example, if \textsc{Magphys} systematically overpredicts the stellar mass when the rest-frame near and mid infrared photometry are poorly constrained \citep{daCunha2015HighZMagphysALESS}.  However, if the measured metallicity is correct (see Appendix \ref{sec:metallicity}), then a lower estimate of the stellar mass (e.g., from a different SED fitting code) will create tension with the MZR (as parameterized by \citealt{Genzel2015DustAndCOscalingRelations}). In other words, the \citetalias{Tacconi2018PHIBSS2MainSequence} gas fraction relation favors a lower stellar mass whereas the MZR favors a higher stellar mass.

Another possibility to reconcile the measured gas fraction with \citetalias{Tacconi2018PHIBSS2MainSequence} is that magnification factor is systematically over-estimated: lower values of $\mu$ will displace points horizontally in the left panel of Fig. \ref{fig:moneyplot}, since ratios of flux-dependent quantities cancel out the factor $\mu$ (under the assumption that the differential lensing effect is not too severe). But in the gas fraction versus stellar mass diagram the horizontal offsets from the \citetalias{Tacconi2018PHIBSS2MainSequence} relation are at least $\sim 1$ dex larger than the vertical offsets (listed in section \ref{sec:scaling_relations}), so magnification alone cannot explain them. Even if we used the image plane uncorrected values (i.e., $\mu=1$), the square symbols in Fig. \ref{fig:moneyplot} would remain below the \citetalias{Tacconi2018PHIBSS2MainSequence} relations.

However, in the high magnification regime, differential lensing can become significant. There are at least three geometrical conditions that can produce a higher magnification for the stellar component relative to the gas component, thus lowering the derived gas fraction: (1) if the UV emission from stars is spatially offset from the molecular gas reservoirs and closer to the lensing caustic lines; (2) if the stars' brightness distribution is more compact than  the gas distribution; or (3), with a combination of the previous two effects. The first condition is certainly possible, since spatial offsets between gas and stars are commonly observed in resolved studies of unlensed, high-redshift galaxies \citep[][and references therein]{Hodge&daCunha2020Review}. For example, \citet{Chen2015ALMAeCDFSMorph} found an average intrinsic scatter in the positional offsets between ALMA \submmw~detections and their HST counterparts of $\approx\ang{;;0.3}$, in a sample of 48 SMGs at $z=1-3$. Similar offsets are found for other gas tracers such as CO lines \citep[e.g.,][]{CalistroRivera2018CoAndDustDistributionALESS}. But for any given system, the actual configuration of gas and stars is random and independent of the foreground lens. Even if we are selecting sources with a highly magnified stellar component, there is nothing in our selection function that favors a less magnified gas component. This means that, statistically, the magnification bias due to spatial offsets in the source plane plays in both directions and eventually cancels out. Then, based on spatial offsets alone, it is unlikely that both our lensed sources and the ones from \citetalias{Saintonge2013MolGasLensedGalaxies} and \citetalias{DessaugesZavadsky2015MolGasLensedSFGs} have a systematically more magnified stellar component. Most importantly, large offsets between unobscured UV emission and dust or gas tracer are preferentially seen in galaxies with high dust obscuration such as in GN20 \citep{Hodge2015ResolvedGN20}, but less commonly in low metallicity galaxies. Regarding the sizes of each component, observational evidence suggest that the background population will have a more compact dust component than the stellar component. This is true for both massive galaxies and intermediate mass galaxies \citep[e.g.,][]{Fujimoto2017DancingALMA1, Kaasinen2020DustSizeAndKinematicsThreeASPECSgxs}. So if this was the case for the galaxies in our sample, it will drive larger magnifications for the gas component pushing the gas fraction to higher values, not lower. In conclusion, it is still possible to have a geometrical configuration of the source system in which lensing preferentially boosts the stellar component over the gas component, for example, with an extended gas distribution relative to stars coupled with a large spatial offset, but such configurations are extremely rare, making it unlikely to produce a systematic effect.  Unfortunately, we cannot currently provide a quantitative estimation of the bias in the magnification. That will require matching HST resolution with ALMA observations of the gas/dust in the arcs.

Now, we consider the case where SFR is overestimated but M\textsubscript{stars} is not. Then, at fixed M\textsubscript{mol}, the depletion timescale appears lower than it actually is. Also, for a fixed stellar mass and redshift,  \citetalias{Tacconi2018PHIBSS2MainSequence} predicts higher gas fractions at higher SFR. Again, this effect can be due to \textsc{Magphys} using priors that prefer templates with higher specific SFR (SFR/M\textsubscript{stars}). But the best fit \textsc{Magphys} models are already on or slightly below the \citetalias{Tacconi2018PHIBSS2MainSequence} MS, so if the stellar masses are correct, then a lower value for the SFR will displace them even further from the MS, toward the locus of quenched galaxies. This is again an implausible scenario, for these galaxies have strong indications of star formation activity (e.g., young ages, blue colors, nebular emission lines, etc.).

Moreover, independent measures of stellar mass and SFR for some of the lensed galaxies in this work have been reported in the literature. For example, \citetalias{Saintonge2013MolGasLensedGalaxies} found SGASJ1226-A to have $\log(\mu M_\text{stars}/\text{M}_\odot)=11.32\pm0.15$ and $\log(\mu\text{SFR}/(\text{M}_\odot\,\text{yr}^{-1}))= 3.15\pm0.08$, which convert to $9.38\pm0.15$ and $1.2\pm0.08$ respectively when matched to our fiducial magnification $\mu=87$. While the SFR reported here is three times lower, the stellar mass is in excellent agreement. More recently, \citet{Vanzella2021SunburstArcClusterFormationEfficiency} reported $\log(M_\text{stars}/\text{M}_\odot)\approx 9.0$ and $\log(\text{SFR}/(\text{M}_\odot\,\text{yr}^{-1}))\approx 1$ for the \textit{Sunburst} galaxy, also consistent with our estimates within the uncertainties. The considerations above plus the agreement with values obtained by other teams suggest that the SED-derived quantities alone cannot account for the tension with  \citetalias{Tacconi2018PHIBSS2MainSequence} relations, thus disfavoring the hypothesis that they hold true in this region of the parameter space.

\subsection{Low metallicity as the driver of the apparent gas deficit} \label{sec:co_deficit}
If the cold gas deficit is not caused by a systematic overestimation of M\textsubscript{stars} and SFR, then the conclusion is that the molecular gas mass might be underestimated. Before exploring this idea, we first note that the gas deficit is observed not only in the four galaxies in our sample, but also in the compilation of lensed sources from \citetalias{Saintonge2013MolGasLensedGalaxies} and \citetalias{DessaugesZavadsky2015MolGasLensedSFGs}. Our results provide additional evidence for a divergence from standard scaling relations in intermediate- to low-mass SFGs at $z\gtrsim 2$. But this result is not exclusive to strongly lensed galaxies. For example,  \citet{Coogan2019SuppressedCOinLowMetGals} put stringent upper limits on the molecular gas mass of five $z\sim 2$ Lyman Break Galaxies (LBGs), with values that also challenge the scaling relations. More recently, \citet{Boogaard2021ColdGasLowMassSFGs} used data from the ASPECS Large Program \citep[e.g.,][]{Aravena2019ASPECSMolGasEvolution} to constrain the gas content of 24 MUSE-selected SFGs at $z=3-4$ in the Hubble Ultra Deep Field (HUDF). At the standard Galactic value for $\alpha_\text{CO}$, \citet{Boogaard2021ColdGasLowMassSFGs} also obtain gas fractions upper limits which are in tension with scaling relations from \citetalias{Tacconi2018PHIBSS2MainSequence} and \citet{Liu2019A3COSMOScoldGasEvolution}.

In both lensed and unlensed studies, metallicity is often pointed out as the main driver of these discrepancies. In particular, it seems that the effect of low metallicity has a stronger impact in the tracer-to-gas conversion factor than expected. In other words, the lack of CO, dust or neutral carbon emission in high-$z$ galaxies may reflect a redshift evolution of $\alpha_\text{CO}$, $\delta_\text{GDR}$, and $X_{[\ion{C}{i}]}$ , respectively. For example, at a fixed metallicity, $\alpha_\text{CO}$ in a high-$z$ SFG should be 5 to 30 times larger than in a local SFG in order to explain the deficit.
Such high values of $\alpha_\text{CO}$ can be produced by CO-faint gas: a decreased metal abundance implies fewer C and O atoms but also a lower proportion of dust grains (high $\delta_\text{GDR}$), thus providing less shielding from the far-ultraviolet radiation that causes CO dissociation. As H\textsubscript{2} is less impacted by this effect \citep[e.g.,][]{GnedinAndDraine2014SelfShieldingMolecularHydrogen}, the abundance of CO relative to H\textsubscript{2} becomes much lower \citep{Bolatto2013COtoH2}. In the Milky Way, a significant amount of H\textsubscript{2} resides outside the CO-bright cores of molecular clouds, as inferred from independent tracers such as $\gamma$-rays \citep{Grenier2005DarkGasSolarNeighborhood}. But in low-metallicity star-forming local dwarf galaxies the CO-faint gas is much more pervasive, filling the regions where most of the carbon is occupying its first ionized state.  In such cases, CO emission cannot trace the bulk of the molecular gas budget and the [\ion{C}{ii}] \SI{158}{\micro\meter} line emerges as an alternative tracer \citep[e.g.,][]{Wolfire2010DarkMolGas, GloverAndClark2012StarFormationInMetalPoorGas, Schruba2012LowCOLuminosityDwarfs, Amorin2016COtoH2, Madden2020TotalMolecularGasDark}.

In principle, $\alpha_\text{CO}(Z)$ should account for the CO-faint gas at low metallicity, but several prescriptions exist in the literature and there is no consensus on the slope of the relation. While most of the published recipes agree that $\alpha_\text{CO, MW} \approx 4.5\,\si{\msun}(\si{\kelvin\kilo\meter\per\second\parsec\squared})^{-1}$ at solar metallicity, the dispersion is very large at lower metallicities. The power-law index $\gamma$ can vary from $-0.65$ \citep{Narayanan2010FormationOfSMGs} to $-3.39$ \citep{Madden2020TotalMolecularGasDark}, but only the steeper laws ($\gamma\lesssim -2.0$,  normalized to $\alpha_\text{CO}(Z_\odot)=\alpha_\text{CO, MW}$) are able to boost M\textsubscript{mol} up to $\sim 1.2$ dex at the metallicity range studied here. Using such recipes \citep[e.g.,][]{Schruba2012LowCOLuminosityDwarfs, Madden2020TotalMolecularGasDark} could put our galaxies closer to the scaling relations found for the gas fraction and depletion time, especially for the SGASJ1226 system which has the lowest metallicities.

A similar argument can be made regarding $\delta_\text{GDR}$. In both observations of local galaxies and predictions from ISM models  $\delta_\text{GDR}$ is found to increase inversely proportional to Z, with a power-law index close to unity  \citep[$\gamma_\text{GDR}\approx-0.85$;][]{Leroy2011COtoH2, Magdis2012EvolvingISMatRedshiftTwo, Sandstrom2013AlphaCOandGDRLocal}.
Other authors find that the scaling steepens at lower metallicity and thus a double power-law is best suited \citep{Remy-Ruyer2014GasToDustLocalGalaxies}. At higher redshift, studies of low mass galaxies have shown that either a shift in normalization \citepalias{Saintonge2013MolGasLensedGalaxies, DessaugesZavadsky2015MolGasLensedSFGs} or a steeper relation is needed to account for the low dust luminosity observed \citep{Coogan2019SuppressedCOinLowMetGals}.

Finally, large uncertainties also affect the determination of the neutral carbon abundance $X_{[\ion{C}{i}]}$ calibration. On one hand, theoretical prescriptions are strongly sensitive to modeling choices, such as the physics of cosmic rays and molecular cloud evolutionary states \citep[][and references therein]{Hodge&daCunha2020Review}. On the other hand, fully empirical calibrations are not yet available. ALMA is starting to fill the gap with observations of [\ion{C}{i}] emission in the local \citep[e.g.,][]{Crocker2019ResolvedAtomicCarbonLocal} and distant \citep[e.g.,][]{Valentino2018SurveyOfAtomicCarbon} Universe, but those results are still dependent on CO or dust-based estimates of M\textsubscript{mol}, which suffer from the aforementioned uncertainties. Here we have used a recipe that was calibrated independently from CO, but  has some other caveats: \citet{Heintz&Watson2020CItoH2} used a small sample of quasar and gamma ray burst absorbers with abundance measurements of both \ion{C}{i}\textsuperscript{*} and H\textsubscript{2}. Absorption lines yield line-of-sight column densities, rather than surface densities, so the authors assume that the ratio between column densities of \ion{C}{i} and H\textsubscript{2} is equal to the surface density ratio. Furthermore, a single line of sight does not necessarily probes all the phases and conditions where neutral carbon and H\textsubscript{2} coexist. Metallicities are also derived in absorption, hence are not directly comparable to the standard nebular emission line determination of oxygen abundances.

Given the large dispersion in tracer-to-gas calibrations in the subsolar metallicity regime and the small size of our sample, it remains difficult to falsify the hypothesis proposed above. In other words, the molecular gas scaling relations of massive galaxies at high redshift may not apply to lower mass galaxies. For example, one could envision a scenario in which recent starburst consumed or dispersed most of the gas available. But this claim cannot be confirmed without a more systematic calibration of gas tracers. Deciding whether the gas deficit in star-forming galaxies with $\text{M}_\text{stars} \lesssim \SI{e10}{\msun}$ is real will require larger samples and deeper integration times using the full ALMA array, even with the aid of strong lensing.

\section{Summary and conclusions} \label{sec:conclusions}

In this paper, we have reported ACA observations of the molecular gas on four star-forming systems at $z\gtrsim 2.5$ that are strongly-lensed by foreground clusters and were selected as bright giant arcs in the optical. We used different methods and tracers, in order to assess possible systematic effects. The resulting detections and upper limits, in combination with ancillary multiwavelength data, allowed us to characterize the global properties of the cold ISM such as gas fraction and gas depletion timescale in these galaxies compared to existing scaling relations. 

 Out of four galaxies of similar masses and SFR, only the most massive, SGASJ0033, was detected in CO(4-3), \neutralc~and dust continuum emission (see Fig. \ref{fig:coline0033}). The galaxy exhibits a very narrow (FWHM $\approx \SI{85}{\kilo\meter\per\second}$) line profile coupled with a low significance asymmetric wing at high velocity. The highly magnified SGASJ1226-A was not detected in either CO(5-4) nor dust continuum. Instead, a companion lensed galaxy at the same redshift, SGASJ1226-B, was detected in Band 7 continuum (see Fig. \ref{fig:rb7_continua}) implying larger amounts of gas than in the A component. Also, no detections were made toward the extremely bright giant arc known as the \sbarc. However, due to the high magnification of the source, the intrinsic upper limits can probe down to $\SI{e9}{\msun}$ in molecular gas mass (assuming the typical tracer-to-gas calibrations, see Fig. \ref{fig:tracer_comparison}). Remarkably, this result implies a gas depletion timescale shorter than $\sim\SI{70}{\mega\year}$.
 
 The inferred gas fraction in the sample is higher than in $z=0$ SFGs, in agreement with the trend of increasing gas fraction at high redshift. However, the gas fraction is roughly $0.5$ to $1.0$ dex lower than predicted by the \citet{Tacconi2018PHIBSS2MainSequence} (\citetalias{Tacconi2018PHIBSS2MainSequence}) scaling relations based on stellar mass, redshift and offset from the MS (see Fig. \ref{fig:moneyplot}). Similarly, gas depletion timescale is also below the expected value based in the \citetalias{Tacconi2018PHIBSS2MainSequence} relation but with a milder offset. The lensed galaxies studied here lie only $\sim 0.6$ to $\sim 0.8$ dex below the locus of more massive unlensed galaxies at the same redshifts (see Fig. \ref{fig:moneyplot}).
 
 To investigate whether the apparent gas deficit is real, we explored systematic offsets that could be driving it. We find that the result is not strongly affected by systematic effects on SED fitting, as it would take unrealistic offsets in M\textsubscript{stars} and SFR to drive the measured discrepancies with the scaling relations. We also explore the effect of differential lensing, but we conclude that it is unlikely to drive a large systematic offset in our sample of lensed galaxies plus the ones from literature. We propose that the apparent gas deficit is rather due to systematic uncertainties in the tracer-to-gas conversion factor dependence on metallicity. Our results favor a scenario in which the emission of CO, dust continuum, and [\ion{C}{i}] are more strongly suppressed than what typical calibrations predict. Of course, this does not rule out the possibility of these galaxies actually have less gas than expected, due for example, to recent starburst episodes having consumed most of the available gas. 

Finally, we stress that in using shallow ACA observations in combination with the lensing effect of massive clusters, it is possible to detect the molecular gas in a $\text{M}_\text{stars}\sim\SI{e10}{\msun}$ galaxy and to put constraints on the molecular gas content of less massive galaxies. Expanding this technique to a statistical sample of giant arcs will tell if the apparent gas deficit is a common feature in lower mass galaxies at high-redshift. However, in order to get accurate estimates of the molecular gas mass, it is necessary to build improved calibrations of the known tracers with respect to metallicity. 

\begin{acknowledgements}
    We thank the anonymous referee for the constructive feedback and helpful comments. We also thank  Ian Smail for the feedback about the manuscript. This paper makes use of the following ALMA 7m array data: ADS/JAO.ALMA\#2018.1.01142.S. ALMA is a partnership of ESO (representing its member states), NSF (USA) and NINS (Japan), together with NRC (Canada), MOST and ASIAA (Taiwan), and KASI (Republic of Korea), in cooperation with the Republic of Chile. The Joint ALMA Observatory is operated by ESO, AUI/NRAO and NAOJ.
    This research is also based on observations made with the NASA/ESA Hubble Space Telescope obtained from the Space Telescope Science Institute, which is operated by the Association of Universities for Research in Astronomy, Inc., under NASA contract NAS 5–26555. These observations are associated with programs 15377, 15101, 15378, 12368 and 14170. In addition, this paper has made use of \textit{Spitzer} Space Telescope data from the NASA/IPAC Infrared Science Archive, which is funded by the National Aeronautics and Space Administration and operated by the California Institute of Technology. SL was funded by project FONDECYT 1191232. LFB was partially supported by CONICYT Project BASAL AFB-170002. MA acknowledges support from FONDECYT grant 1211951, CONICYT + PCI + INSTITUTO MAX PLANCK DE ASTRONOMIA MPG190030 and CONICYT+PCI+REDES 190194.
    This research made use of Astropy,\footnote{\url{http://www.astropy.org}} a community-developed core Python package for Astronomy \citep{Astropy2013PaperI, Astropy2018PaperII}. All figures were prepared using Matplotlib \citep{Hunter2007MatplotlibPaper} and Seaborn \citep{Waskom2020Seaborn}. The scaling relations used in this paper were obtained from convenience routines included in the A\textsuperscript{3}COSMOS gas evolution library \footnote{\url{https://ascl.net/1910.003}} \citep{Liu2019A3COSMOScoldGasEvolution}.
\end{acknowledgements}

%
\bibliographystyle{aa} 
\bibliography{main} 
%
\appendix
\section{Stellar $Z$ as a proxy of gas-phase metallicity} \label{sec:metallicity}
The scaling relations for $\alpha_\text{CO}$ or $\delta_\text{GDR}$ are calibrated against gas-phase metallicity (parameterized as the oxygen abundance, $12 + \log(\text{O/H})$), which is obtained via rest-frame optical nebular line ratio indicators such as N2, O3N2 ([N II]$\lambda 6584$/H$\alpha$ and [O III]$\lambda 5007$/H$\beta$/N2 respectively, \citealt{PettiniAndPagel2004Metallicity}) or $R_{23}$ (([O II]$\lambda 3727$ + [O III]$\lambda\lambda 4959,5007$)/H$\beta$, \citealt{McGaugh1991MetallicityR23}). At the redshifts considered here, these lines can only be accessed through near-infrared bands. In the case of SGASJ1226, H$\alpha$ is shifted outside the atmospheric NIR window and hence can only be observed from space. Nevertheless, \citet{Wuyts2012StellarPopsLensed} presented a Keck NIRSPEC spectrum of SGASJ1226 in which [O II]$\lambda 3727$, [O III]$\lambda\lambda 4959, 5007$ and H$\beta$ were successfully identified and the reported fluxes were later used by \citet{Saintonge2013MolGasLensedGalaxies} to infer $12+\log(\text{O/H})=8.27\pm0.19$ based on the $R_{23}$ index. For the \sbarc, ESO X-Shooter NIR spectra was presented in \citep{Vanzella2020BowenFluoresenceSunburstArc}, but without reporting any metallicity indicator nor individual line fluxes. Finally, as we already mentioned in the text, SGASJ0033 was observed with the SINFONI instrument on the VLT by \citet{Fischer2019outflowJ0033}, who measured the flux of several diagnostic lines but also excluded an estimate of metallicity. In summary, the rest-frame optical spectrum of our sample has already been explored, but the diversity of instruments and the lack of consistent metallicity indicators prevent us from using gas-phase oxygen abundances in the $\alpha_\text{CO}, \delta_\text{GDR}$ or X\textsubscript{[CI]} recipes.

Fortunately, the high quality MegaSaura spectra permitted \citet{Chisholm2019MegaSauraMetallicity} to fit \textsc{Starburst99} single stellar population models and infer ages and stellar metallicities. Here we used the latter as a proxy of gas-phase metallicity, as pointed out in Sect. \ref{sec:gas_mass}, so we can compare three values determined in a uniform way from the same instrument. The rationale behind this choice is that, due to the young age of the UV-bright population accessible in the $\sim$\SIrange{1200}{3000}{\angstrom} rest-frame range, the fitted stellar metallicity is likely equal to the gas-phase metallicity. This hypothesis was already tested by \citet{Chisholm2019MegaSauraMetallicity}, who found a tight agreement between these two metallicities for the subset of MegaSaura galaxies which had both measurements available. To further validate the star-gas metallicity equivalence, we performed an additional test: taking SGASJ0033 as a benchmark, we used the published (narrow) line fluxes to infer oxygen abundance form the N2 and O3N2 indicators and compare to both stellar metallicities reported by \citet{Chisholm2019MegaSauraMetallicity}, namely $Z_\text{SB99}^{\star}$ and $Z_\text{BPASS}^{\star}$.

We obtained the reddening-corrected narrow line fluxes from Table 1 of \citet{Fischer2019outflowJ0033}. These measurements correspond to the central $\ang{;;0.5}\times\ang{;;0.5}$ region of the SGASJ0033 arc, but we do not expect significant AGN contamination since the two-component analysis presented in their Figure 4 predicts the narrow component to be well within the star formation dominated region of the \citet{Baldwin1981BPTpaper} diagram (BPT). We computed the N2 and O3N2 indices and converted them to oxygen abundance using the \citet{PettiniAndPagel2004Metallicity} calibration. The results are shown in Table \ref{tab:metal_j0033}, where all metallicity indicators displayed in the second column are consistent within 0.2 dex. 

\begin{table}[!hbt]
    \centering
    \caption{Different metallicity indicators for SGASJ0033 and their effect in gas tracer conversion factors}
    \label{tab:metal_j0033}
    \begin{tabular}{lcccc}
        \hline
        \hline
         Method & $12 + \log\text{(O/H)}$ & $\alpha_\text{CO}$\tablefootmark{a} & $\delta_\text{GDR}$\tablefootmark{b} &$X_\text{[C I]}^{c}\times10^5$ \\
         \hline
         $Z^{\star}_\text{SB99}$ & $8.59\pm 0.02$\tablefootmark{d} & $4.5\pm 0.1$ & $117 \pm 5$  & \num{1.3+-0.8}\\
         $Z^{\star}_\text{BPASS}$ & $8.54\pm 0.02$\tablefootmark{d} & $4.8 \pm 0.1$ & $129 \pm 5$ &  \num{1.2+-0.7}\\
         N2\textsubscript{PP04} & $8.65 \pm 0.10$\tablefootmark{e} & $4.3 \pm 0.4$ &  $104 \pm 20$ & \num{1.5+-1.0}\\
         O3N2\textsubscript{PP04} & $8.45 \pm 0.10$\tablefootmark{e} & $5.3\pm 0.7$ & $154 \pm 30$ & \num{0.9+-0.6} \\
         MZR\textsubscript{G12} & $8.54\pm0.20$ & $4.8\pm 1.1$ & $129 \pm 40$ &  \num{1.2+-0.9}\\
         \hline
         \hline
    \end{tabular}
    \tablefoot{
        \tablefoottext{a}{CO-to-H\textsubscript{2} conversion factor computed using metallicity-dependent recipe from \citet{Genzel2015DustAndCOscalingRelations}, in units of $\si{\msun}\,(\si{\kelvin\kilo\meter\per\second\parsec\squared})^{-1}$.}
        \tablefoottext{b}{Gas-to-dust mass ratio computed using the metallicity-dependent recipe from \citetalias{Tacconi2018PHIBSS2MainSequence} as explained in Sect. \ref{sec:gas_mass}.}
        \tablefoottext{c}{Recipe from \citet{Heintz&Watson2020CItoH2}, in units of $\si{\msun}\,(\si{\kelvin\kilo\meter\per\second\parsec\squared})^{-1}$.}
        \tablefoottext{d}{Taken from \citet{Chisholm2019MegaSauraMetallicity}.}
        \tablefoottext{e}{Based on the narrow line fluxes reported by \citet{Fischer2019outflowJ0033}.}
    }

\end{table}

\end{document}

%% file: summary_targets.tex
\begin{tabular}{ccccccccc}
    \hline
    \hline
    \multirow{2}{*}{Target} & Coordinates (ICRS) & \multirow{2}{*}{$z^\text{ref}$} & \multirow{2}{*}{Discovery paper}  & B4 RMS\tablefootmark{a} & B7 RMS\tablefootmark{b} &  \multicolumn{2}{c}{Resolution\tablefootmark{c}} \\
    & RA, Dec & & & \si{\milli\jansky\kilo\meter\per\second} & \si{\micro\jansky} &  B4 & B7 \\
    \hline
     \sbarc & 15:49:59.7, -78:11:13.6 & 2.371\textsuperscript{1} &  \citet{Dahle2016SunburstArcDiscovery} 
     &  16.8 & 245 & \ang{;;11.1} & \ang{;;4.9}  \\
     SGASJ1226 & 12:26:51.3, +21:52:19.8 & 2.923\textsuperscript{2} &  \citet{Koester2010ArcJ1226Discovery}  & 26.4 & 170 & \ang{;;9.4} & \ang{;;4.3}  \\
     SGASJ0033 & 00:33:41.6, +02:42:17.4 & 2.388\textsuperscript{2} &  \citet{Rigby2018MegaSauraI} & 16.8 & 170  & \ang{;;9.5} & \ang{;;3.6} \\
     \hline
     \hline
\end{tabular}

%% file: table870um_detections.tex
\begin{tabular}{rcc}
  \hline
  \hline
 Source &  $S_\nu(\SI{870}{\micro\meter})$ & $S_\nu(\SI{2.1}{\milli\meter})$ \\

   & \si{\milli\jansky}& \si{\milli\jansky} \\
  \hline 
                  SGASJ1226-A.1   &   $<\num{0.51}$ & $<\num{0.28}$  \\
                  SGASJ1226-A.2   &  $<\num{1.0}$ & $<\num{0.56}$ \\
                  SGASJ1226-B.1   &  \num{1.12+-0.30} & $<\num{0.28}$  \\
                  SGASJ0033-A.1   &  \num{4.0+-0.4} & - \\
                  SGASJ0033-A.2   &  \num{0.9+-0.3} & - \\
                            A.1+A.2\tablefootmark{a} & \num{4.9+-0.5} & \num{0.37+-0.10} \\
                  \sbarc-S2     &  $<\num{0.83}$ & $<\num{0.24}$ \\ 
                  \hline
                  \multicolumn{3}{l}{\textit{Nontarget detections}} \\
                  \hline
                  SGASJ1226-G1    &  \num{1.20+-0.33} & $<\num{0.28}$\\
                  SGASJ0033-G1    & \num{5.0+-0.9} & $<\num{0.35}$\\
                  SGASJ0033-SMG   &  \num{5.1+-1.1} & $<\num{0.35}$ \\
  \hline
  \hline
\end{tabular}

%% file: sed_results.tex
\begin{tabular}{cccccc|cc}
    \hline
    \hline
    Name & $\mu$ & $\log(\text{SFR}_{100})$ & $\log(M_\text{stars})$ & $\log(
    L_\text{dust})$ & $\log\left(M^{\text{SED}}_\text{dust}\right)$ & $\log\left(M^{\submmw}_\text{dust}\right)$ & $\log\left(M^{\SI{2.1}{\milli\meter}}_\text{dust}\right)$\\
    \hline
	     & &  \si{\msun\per\year} & \si{\msun} & \si{\lumsol} & \si{\msun} & \si{\msun} & \si{\msun} \\
	\hline
    \sbarc & 171 & $1.20_{-0.13}^{+0.07}$ & $ 9.44_{-0.11}^{+0.17}$ & $10.98_{-0.30}^{+0.18}$ & $<7.1 $ & $<6.8$ & $<7.0$\\
    SGASJ1226-A\tablefootmark{a} & 87 & $0.65_{-0.05}^{+0.11}$ & $9.38_{-0.11}^{+0.11}$ & $<10.7$ & $<6.4$ & $<6.5$ & $<7.4$ \\
    SGASJ1226-B & 30 & $1.12_{-0.08}^{+0.21}$ & $9.73_{-0.16}^{+0.17}$ & $10.87_{-0.12}^{+0.37}$ & $6.8_{-0.2}^{+0.2}$ & $7.3\pm0.1$ & $<7.8$\\
    SGASJ0033-A\tablefootmark{a} & 67 & $1.12_{-0.19}^{+0.22}$ & $10.10_{-0.17}^{+0.16}$ & $11.12_{-0.24}^{+0.25}$ & $7.10_{-0.14}^{+0.19}$ & $7.59 \pm 0.04$ & $7.65 \pm 0.14$\\
    \hline
    SGASJ1226-G1\tablefootmark{b} & 2.7 &  $1.20_{-0.20}^{+0.17}$ & $10.09_{-0.08}^{+0.08}$ & $11.24_{-0.18}^{+0.15}$ & $8.01_{-0.15}^{+0.17}$ & $8.22 \pm 0.12$ & $<8.9$\\
    SGASJ0033-G1\tablefootmark{c} & 5.4 & $1.81_{-0.33}^{+0.34}$ & $10.55_{-0.24}^{+0.27}$ & $11.89_{-0.32}^{+0.33}$ & $8.19_{-0.16}^{+0.18}$ & $8.63 \pm 0.08$ & $<8.8$\\
    \hline
    \hline
\end{tabular}